\documentclass[journal]{IEEEtranTCOM}
\normalsize
\usepackage[pdftex]{graphicx}
  \usepackage{epstopdf}
  \DeclareGraphicsExtensions{.pdf,.jpeg,.png,.eps}
\usepackage{color,balance}
\usepackage[cmex10]{amsmath}
\usepackage{amssymb}
\usepackage{algorithmic}
\usepackage{array}
\usepackage{mdwmath}
\usepackage{amsthm}
\usepackage{multirow}
\usepackage{booktabs}
\usepackage{comment}
\usepackage{graphicx}
\usepackage{textgreek}
\usepackage{float}
\usepackage{cite}

\usepackage{mdwtab}
\usepackage{tabularx}
\usepackage{balance}
\usepackage{cite}
\makeatother

\allowdisplaybreaks
\newtheoremstyle{mystyle}
  {3pt}
  {3pt}
  {\itshape} 
  {\parindent}
  {\bfseries}
  {\upshape{:}}
  {.5em}
  {}

\usepackage{color,soul}
\usepackage{setspace}
\theoremstyle{mystyle}

\theoremstyle{mystyle}  

\theoremstyle{mystyle}

\usepackage[british,UKenglish]{babel}
\usepackage{subcaption}
\usepackage[font={small}]{caption}
\graphicspath{{./figures/}}

\begin{document}
%\linespread{0.9}

\title{Edge-Native Intelligence for 6G Communications Driven by Federated Learning: A Survey of Trends and Challenges }

\author{Mohammad Al-Quraan,~\IEEEmembership{Graduate Student Member,~IEEE,} Lina Mohjazi,~\IEEEmembership{Senior Member,~IEEE,} Lina Bariah,~\IEEEmembership{Senior Member,~IEEE,} Anthony Centeno, Ahmed Zoha,~\IEEEmembership{Senior Member,~IEEE,} Kamran Arshad, Khaled Assaleh, Sami Muhaidat,~\IEEEmembership{Senior Member,~IEEE,} M\'{e}rouane Debbah,~\IEEEmembership{Fellow,~IEEE,} and Muhammad Ali Imran,~\IEEEmembership{Fellow,~IEEE}

\thanks{M. Al-Quraan, L. Mohjazi, A. Centeno, and A. Zoha are with the James Watt School of Engineering, University of Glasgow, Glasgow, G12 8QQ, UK, (e-mail: m.alquraan.1@research.gla.ac.uk, \{Lina.Mohjazi, Anthony.Centeno, Ahmed.Zoha\}@glasgow.ac.uk).}
\thanks{L. Bariah is with the Technology Innovation Institute, 9639 Masdar City, Abu Dhabi, UAE, (e-mail: lina.bariah@ieee.org).}
\thanks{K. Arshad and K. Assaleh are with the Artificial Intelligence Research Center (AIRC), College of Engineering and Information Technology, Ajman University, Ajman, UAE, (e-mail: \{k.arshad, k.assaleh\}@ajman.ac.ae).}
\thanks{S. Muhaidat is with the Center for Cyber-Physical Systems, Department of Electrical Engineering and Computer Science, Khalifa University, Abu Dhabi 127788, UAE, and also with the Department of Systems and Computer Engineering, Carleton University, Ottawa, ON K1S 5B6, Canada, (e-mail: muhaidat@ieee.org).}
\thanks{M. Debbah is with the Technology Innovation Institute, 9639 Masdar City, Abu Dhabi, UAE, (email: merouane.debbah@tii.ae) and also with CentraleSupelec, University Paris-Saclay, 91192 Gif-sur-Yvette, France.}
\thanks{M. A. Imran is with the James Watt School of Engineering, University of Glasgow, Glasgow, G12 8QQ, UK, and also with Artificial Intelligence Research Center (AIRC), College of Engineering and Information Technology, Ajman University, Ajman, UAE, (e-mail: Muhammad.Imran@glasgow.ac.uk)}}

\maketitle

\markboth{}{}
%\vspace*{-1.5cm}
\begin{abstract}
New technological advancements in wireless networks have enlarged the number of connected devices. The unprecedented surge of data volume in wireless systems empowered by artificial intelligence (AI) opens up new horizons for providing ubiquitous data-driven intelligent services. Traditional cloud-centric machine learning (ML)-based services are implemented by centrally collecting datasets and training models. However, this conventional training technique encompasses two challenges: (i) high communication and energy cost and (ii) threatened data privacy. In this article, we introduce a comprehensive survey of the fundamentals and enabling technologies of federated learning (FL), a newly emerging technique coined to bring ML to the edge of wireless networks. Moreover, an extensive study is presented detailing various applications of FL in wireless networks and highlighting their challenges and limitations. The efficacy of FL is further explored with emerging prospective beyond fifth-generation (B5G) and sixth-generation (6G) communication systems. This survey aims to provide an overview of the state-of-the-art FL applications in key wireless technologies that will serve as a foundation to establish a firm understanding of the topic. Lastly, we offer a road forward for future research directions.
\footnote{  Copyright~\copyright~2023 IEEE. Personal use of this material is permitted.  Permission from IEEE must be obtained for all other uses, in any current or future media, including reprinting/republishing this material for advertising or promotional purposes, creating new collective works, for resale or redistribution to servers or lists, or reuse of any copyrighted component of this work in other works.}
\end{abstract}
%\vspace*{-0.5cm}
%\doublespacing
%\vspace*{-0.5cm}

\section{INTRODUCTION}
Recent years have witnessed an unprecedented increase in the number of connected objects, which is attributed to the emergence of novel technological trends as well as the evolution of connected intelligence paradigms, promoting massive scale connectivity \cite{[3.63]}. In specific, the number of internet-of-things (IoT) devices per human was 1.84 in 2010 with a total of 12.5 billion devices, while in 2020, this number increased to 6.58 devices per human with nearly a total of 50 billion devices \cite{[9.30]}. With the remarkable revolutionary advancements in the field of wireless communications, it is envisioned that these numbers will continue to rise exponentially. Accordingly, enlarging connected devices, such as IoT, smartphones, industry machines, etc., will create a bottleneck on the limited resources of wireless networks. Therefore, there will be a continuous need to develop the existing network infrastructure to meet diversified demands.
 
According to the International Telecommunications Union (ITU) and the 3rd Generation Partnership Project (3GPP), the fifth generation (5G) wireless networks are designed to deliver improved quality of experience (QoE) by offering enhanced data rate, reliability, capacity, and energy efficiency. In light of this, 5G wireless systems were mapped out based on three fundamental concepts, namely, enhanced mobile broadband (eMBB), ultra-reliable and low latency communications (URLLC), and massive machine-type communications (mMTC) \cite{[1.2]}. Nevertheless, the rise of services like extended reality and massive IoT, and the expected future applications such as holographic communications and multi-sense experience, impose far more stringent requirements than 5G networks and shed light on the next network improvements. Hence, the research will be shifted towards sixth-generation (6G) communication networks.
 
The fast-growing number of connected devices, coupled with the development of wireless communication infrastructures and the capability of embracing a wide range of intelligent applications, have resulted in unprecedented volumes of produced data traffic that need to be stored and processed; yielding the new concept of big data \cite{[1.3]}. To harness the benefits of this data, artificial intelligence (AI), especially machine learning (ML), has become the cutting-edge technology that has the potential to exploit big data to deliver pervasive smart services and applications \cite{[3.67]}. ML models are trained to perform diversified tasks by exploring hidden data patterns and drawing the value of such data to predict useful outcomes for several use cases, such as medical diagnosis and natural language processing \cite{[11.6]}.
 
In conventional ML algorithms, model training is performed centrally in cloud-based servers\cite{[4.5]}. Datasets are collected and stored in one location, processed, and then employed to train ML models using one or multiple servers. This centralised nature of ML models limits their applicability for several emerging wireless network applications. The limitations include the following:
\begin{itemize}
    \item Increased communication overhead between the end devices and the cloud resulting in network congestion and high energy consumption.
    \item The privacy is not by design, so security and data privacy are always a concern for conventional approaches.
    \item The propagation delay experienced in such ML techniques limits the implementation of centralised learning in real-time applications.
\end{itemize}

In light of this, federated learning (FL) has emerged as a promising solution to tackle the aforementioned challenges of centralised ML \cite{[3.11]}. FL is a collaborative ML algorithm that uses distributed entities' datasets for local model training without the need to exchange any raw data with a central server. In FL, the role of cloud-based servers is limited to aggregating local models to develop a global model to be shared with all nodes in the network. Initially, a centralised server broadcasts initial model parameters to participating nodes, which leverage these parameters and their onboard resource capabilities for local model training. Next, once the training round is finished, each participant will send the local model updates to the FL server to aggregate the received local models. For enhanced accuracy, model training in FL is performed over multiple iterations; hence, after each training round, the server shares updated model parameters with participating devices to be utilised for the next training round. By using FL, the amount of data that needs to be sent to the server is reduced significantly, allowing only model updates to be sent to the server and hence, alleviating the pressure on the network resources. Furthermore, FL protects the endpoints' data privacy and security by allowing model training locally, where the data is generated.

\subsection{Related Surveys in the Literature}
FL has attracted numerous interests and has been implemented in diverse applications across many areas. Notably, several surveys have been published since the advent of the FL algorithm. Table \ref{RefWorks} summarises these surveys and highlights their significance. Here we outline the surveys in chronological order based on the publication date. The work in \cite{[3]} categorises the FL systems into three categories, namely, vertical, horizontal, and federated transfer learning (FTL), and discusses the privacy techniques used in FL. The authors in \cite{[9.14]} highlight the need to implement ML at the wireless network edge to facilitate reliable and low-latency communication. They explore the key building blocks of ML that allow for the transition from centralised, cloud-based model training to decentralised training techniques such as FL. Furthermore, a thorough investigation of the technical and theoretical frameworks of several case studies illustrates the importance of edge intelligence for beyond 5G (B5G) networks. Later, Kairouz \emph{et al.} \cite{[2.1]} introduce recent advances in FL by discussing techniques used to improve FL efficiency, explaining the methods used to preserve user data privacy, and how to make FL algorithms robust against attacks. The authors in \cite{[2.5]} discuss possible threats and attacks, and highlight their implications to future FL algorithms, whereas \cite{[2.6]} discusses FL properties and associated challenges in comparison to traditional distributed data centre computing. Du \emph{et al.} \cite{[2.7]} outline the importance and technical challenges of implementing FL in vehicular IoT networks.
\begin{table}
\fontsize{7}{11}\selectfont
\caption{Summary of Relevant FL Surveys. }
\begin{tabular}{clll}
\hline
\textbf{Ref.} & \begin{tabular}[c]{@{}l@{}} \textbf{Date} \end{tabular}& \textbf{Authors} &\textbf{Article Main Topic}                   \\ \hline
\hline
\cite{[3]}      & Jan. 2019 & Q. Yang, \emph{et al.}                         & \begin{tabular}[c]{@{}l@{}} Categories of FL and Privacy\\ Techniques \end{tabular}       \\ \hline
\cite{[9.14]}   & Oct. 2019 & J. Park, \emph{et al.}                        & \begin{tabular}[c]{@{}l@{}}Edge ML in Beyond 5G\\ Networks\end{tabular}   \\ \hline
\cite{[2.1]}    & Dec. 2019 & P. Kairouz, \emph{et al.}                       & \begin{tabular}[c]{@{}l@{}} FL Advances, Problems and\\ Challenges \end{tabular}        \\ \hline
\cite{[2.5]}    & Mar. 2020 & L. Lyu, H. Yu, Q. Yang                  & FL Threats and   Attacks                      \\ \hline
\cite{[2.6]}    & May 2020 & T. Li, \emph{et al.}                          & FL Implementation   Challenges                \\ \hline
\cite{[2.7]}    & May 2020 & Z. Du, \emph{et al.}                          & \begin{tabular}[c]{@{}l@{}} FL Challenges in   Vehicular\\ Networks \end{tabular}         \\ \hline
\cite{[2.9]}    & July 2020 & M. Aledhari, \emph{et al.}                    & \begin{tabular}[c]{@{}l@{}} FL Protocols and Enabling\\ Technologies \end{tabular}      \\ \hline
\cite{[2.4]}    & July 2020 & \begin{tabular}[c]{@{}l@{}} V. Kulkarni, M. Kulkarni,\\ A. Pant \end{tabular} & FL Personalisation Techniques                \\ \hline
\cite{[2.3]}    & July 2020 & W. Yang, \emph{et al.}                          & FL in Mobile Edge   Networks                  \\ \hline
\cite{[10.2]}    & Dec. 2020 & M. Chen, \emph{et al.}                          & Collaborative FL                  \\ \hline
\cite{[2.11]}   & Jan. 2021 & Q. Li, \emph{et al.}                           & Thorough FL   Categorisation                  \\ \hline
\cite{[2.12]}   & Feb. 2021 & O. A. Wahab, \emph{et al.}                      &  \begin{tabular}[c]{@{}l@{}}FL in Communication   and\\   Networking Systems\end{tabular}\\ \hline
\cite{[2.2]}    & Apr. 2021 & S. Abdulrahman, \emph{et al.}                   & \begin{tabular}[c]{@{}l@{}} FL Architecture Extensive\\ Explanation \end{tabular}       \\ \hline
\cite{[10.1]}    & June 2021 & L. Khan, \emph{et al.}                   & \begin{tabular}[c]{@{}l@{}} FL Integration with IoT \\Networks \end{tabular}       \\ \hline
\cite{[10.3]}    & Dec. 2021 & Z. Yang, \emph{et al.}                   & \begin{tabular}[c]{@{}l@{}} FL Implementation in Wireless \\Communications \end{tabular}       \\ \hline
\cite{[11.1]}    & Mar. 2022 & A. Z. Tan, \emph{et al.}   & Taxonomy of personalised FL       \\ \hline
\cite{[11.2]}    & June 2022 & B. Ghimire, D. B. Rawat   & Cybersecurity and FL in IoT      \\ \hline
\end{tabular}
\label{RefWorks}
\end{table}
 
The contribution in \cite{[2.9]} spots the light on the concept of FL and illustrates some of the enabling technologies and recent research that addresses different FL perspectives. The study in \cite{[2.4]} discusses the implications of training the FL model using heterogeneous datasets, and presents recent research that applies personalisation to overcome the data heterogeneity problem. While \cite{[2.3]} is restricted in presenting the challenges associated with deploying FL in mobile edge networks only and provides the developed solutions that optimise these networks. Reliance on a central controller to organise the FL training process in IoT networks can limit the FL applications, and this issue is the authors' main focus in \cite{[10.2]}. Accordingly, they have proposed a collaborative FL (CFL) framework where clients can implement FL with less dependence on the central server. CFL enables clients to engage in FL directly or indirectly, where some users are directly connected to the server while others are associated with neighbouring clients. Furthermore, this survey presents the original FL's architecture, benefits, and shortcomings compared to CFL. In \cite{[2.11]}, the authors present a thorough categorisation of FL in different aspects and discuss the existing solutions with their limitations in enabling FL. \cite{[2.12]} presents a tutorial on FL technologies and the associated challenges in communication and networking systems. The survey in \cite{[2.2]} explains the FL architecture, system model and design, application areas, privacy and security, and resource management. The work in \cite{[10.1]} presents a new taxonomy of FL in the context of IoT networks and explores FL's recent developments toward enabling intelligent IoT applications. Moreover, this survey introduces a set of metrics that can be considered when evaluating the performance of new FL algorithms. The review paper \cite{[10.3]} highlights the requirements for FL in wireless communications, particularly for envisioned 6G systems. Besides, the motivation for using FL and the main obstacles accompanying FL implementation are discussed. \cite{[11.1]} introduces the key motivation and the taxonomy of personalised FL, which is the technique used to handle the statistical heterogeneity of real-world datasets to learn ML models collaboratively. The authors in \cite{[11.2]} study the use of FL in cybersecurity and vice versa, and discuss several approaches that address IoT networks' performance issues when deploying FL.

\subsection{Contributions}
It is worth emphasising that to the best of our knowledge, no prior works presented a comprehensive study of FL potentials and applications for various existing/next-generation wireless networks. Besides, most of the aforementioned survey papers generally focus on specific technological trends or aspects associated with FL applications. Conversely, in this survey, we provide a systematic review with a featured presentation that leads the reader to a thorough understanding of the FL algorithm and its recent advents, as well as its envisioned implementations in various types of B5G/6G wireless networks. Moreover, this article offers numerous future research opportunities derived from the latest trending technologies that have not been covered by any previous surveys to the best of the authors' knowledge.
The following points demonstrate our main contributions:
\begin{itemize}
    \item We present a concrete conceptual background on the working principles of the FL algorithm. Also, we describe its architecture, categories, operation, and optimisation schemes.
    \item We explore the enabling technologies that create the stepping stones for facilitating the operation of FL.
    \item We provide an in-depth discussion of the key drivers for deploying FL in state-of-the-art wireless applications, taking into account the associated performance metrics and ongoing research. Moreover, we discuss the vision for integrating FL with new potential prospective areas in future wireless networks.
    \item The survey delves into highlighting the challenges associated with the operation of FL in emerging wireless technologies, and identifies the approaches proposed to tackle them.
    \item We offer a look ahead towards unexplored possibilities drawn from modern technology trends to reap the benefits of FL implementations in the context of cutting-edge future research directions.
\end{itemize}
It is noteworthy that the survey structure is organised and written in a distinct taxonomy to make it easier for the reader to navigate and recognise the contributions made in each area. 

\begin{figure*}
\centering
 \includegraphics[width=0.9\textwidth]{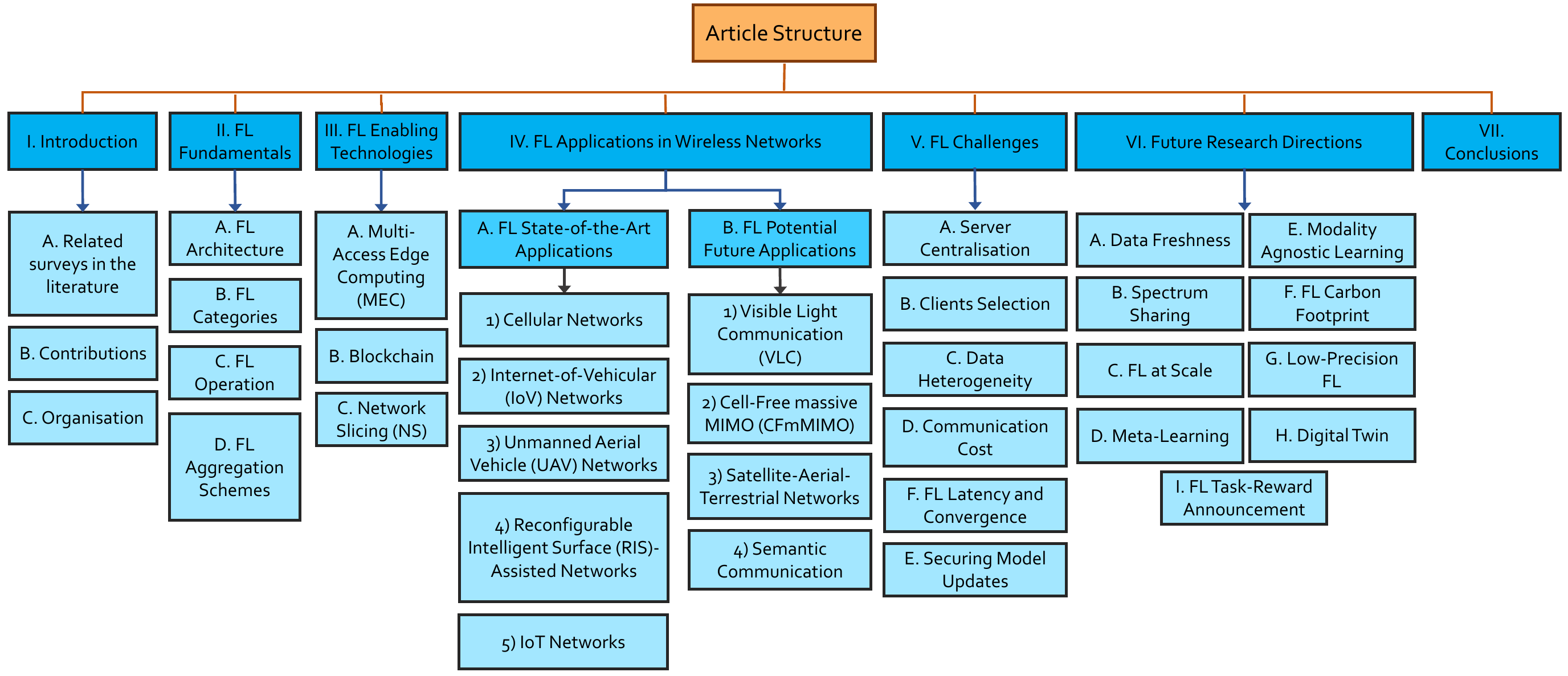}
 \caption{Illustrative diagram of the paper structure.}
 \label{whole_structure}
\end{figure*}

\subsection{Organisation}
The rest of this paper is organised as follows. Section \ref{fundementals} introduces the fundamental aspects of the FL framework covering architecture, categories, operation, and aggregation schemes. Then, we give the key enabling technologies of FL in Section \ref{enabling tech.}. Section \ref{Apps} presents a comprehensive study of FL applications in various wireless networks. In addition, this section discusses its applicability in new potential areas of B5G/6G networks. After that, the FL challenges and corresponding mitigating techniques are outlined in Section \ref{Challenges}. Section \ref{Discussions} points out future research directions from different perspectives. Finally, Section \ref{Conclusion} gives concluding remarks. Fig. \ref{whole_structure} illustrates the detailed outline of this survey.

\section{PRELIMINARY: FL Fundamentals} \label{fundementals}
The concept of FL has attracted significant attention in academia and industry \cite{[3.3]}. The key principle of FL is to construct a generalised global model by performing distributed model training. The recent advancement in edge devices' communication and computation capabilities and the large amount of data generated and stored locally on the devices facilitate the spread of this emerging technology widely. This section presents the fundamentals, architecture, categories, operation principles, and aggregation schemes of FL algorithms.

\subsection{FL Architecture}
\begin{figure}
\centering
 \includegraphics[width=0.45\textwidth]{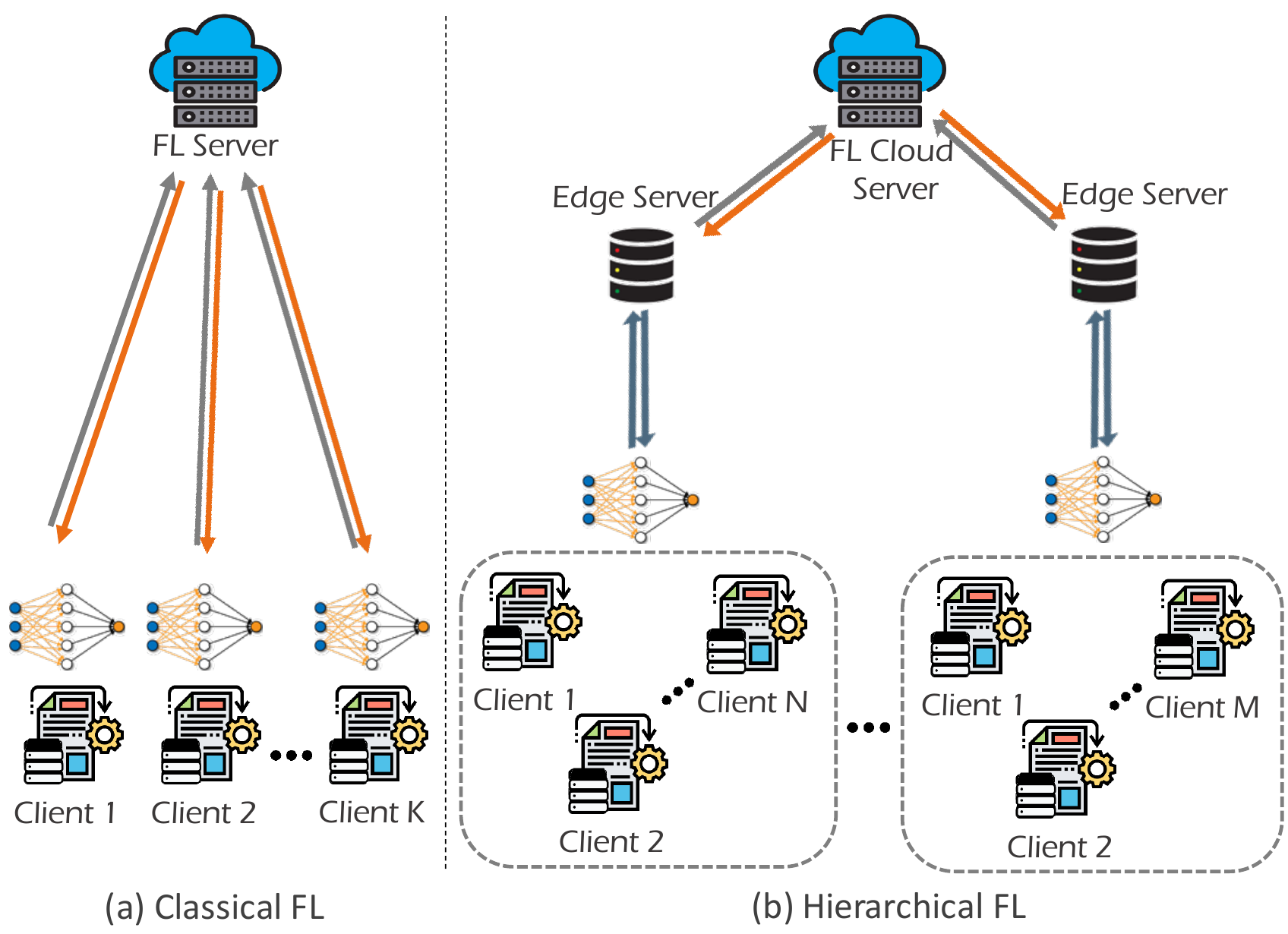}
 \caption{Types of FL architecture (a) Classical FL in client-server architecture (b) Hierarchical FL in client-edge-server architecture.}
 \label{fl_architecture}
\end{figure}

Based on the nature of the network, the architecture of FL can be categorised into classical and hierarchical FL (HFL). The classical FL approach consists of two main parts: the server and the participating clients \cite{[3.11]}, as illustrated in Fig. \ref{fl_architecture}(a). The FL server must have certain specifications to orchestrate the FL process efficiently. These specifications are drawn from the considered ML technique and the number of clients. For instance, training a deep learning (DL) model through many clients requires a high server capacity, a huge computation capability, high-speed interfaces, and locating the server in close proximity to the clients. On the other hand, the specifications may be less stringent when considering simpler models of neural networks and a few clients. At the beginning of the FL process, the server will initiate the training procedure by sharing a new or pretrained model with the participating clients. After that, the clients will personalise the received model by training it based on their local data, and then share their local models with the server for aggregation and global model update. On the other hand, HFL framework \cite{[48]}, depicted in Fig. 2(b), optimally fits in heterogeneous networks that include different cell coverage. This architecture is introduced to alleviate the bandwidth (BW) overhead at the FL servers, resulting from the large number of model updates communicated from the clients. Furthermore, HFL can reduce the communication latency experienced between the clients and the server by reducing the link distance. The HFL framework consists of two stages; in the first one, the clients send and receive the model parameters by communicating with a server located at the small base station (SBS), i.e., the edge server, and the server performs local model aggregation. Meanwhile, in the second stage, the edge servers send the aggregated models to a central server that can be located at the macro base station (MBS) or in the cloud, in which the server performs edge model aggregation for global model update and sends it back to the edge servers.

It is worth mentioning that the need for robust communications between the clients and the FL server, mainly when the latter is located in the cloud, is mandatory to guarantee a seamless FL training process. However, the current internet links' capacity is insufficient to meet the emergency demands, along with the growing connectivity needs from different sectors, such as industry, education, and transportation. This results in the need to move to the new concept of worldwide decentralised internet, which can be achieved using decentralised mesh networks \cite{[9.1]}. Such networks rely on establishing connections between different nodes, i.e., consumers and businesses, to make alternative ways of connectivity other than the known centralised internet service provider connection. Decentralised mesh networks are reliable for maintaining the connections between the participating clients and the FL server and ensuring a smooth training process.

\subsection{FL Categories}
Given the significant role of local datasets in realising efficient training and assuming the data is maintained in a 2D matrix form, rows represent data samples, and columns indicate features. FL systems can be classified based on the data distribution characteristics between different parties into horizontal, vertical, and FTL \cite{[2.11]}.

\vspace{0.2cm}
\subsubsection{\textbf{Horizontal FL}}
This is the most common category of FL, also called sample-based FL. The unique characteristic of this category is that the datasets of different parties share the same feature space while differing in the sample space. For example, two regional educational institutes have similar interests in monitoring the research outcomes, representing the feature space, while they have different research groups that denote the sample space. This category facilitates the adoption of a unified ML model with the same architecture for all datasets. Therefore, the global model can be obtained by averaging all local updates. FedAvg technique \cite{[1.5]} is an example of this type of FL system.

\vspace{0.2cm}
\subsubsection{\textbf{Vertical FL}}
This category, referred to as feature-based FL, can be exploited when two or more datasets share the same sample space while their feature spaces are distinct. For instance, considering two different parties in the same city, one is a healthcare institute, whereas the other is an e-commerce company that records the customers' buying habits. Their user sets are most likely to have residents from that area, which means the same sample space. The objective here is to exploit the different features of these two parties to build a model that predicts the future health status of the residents based on their buying practices. When implementing vertical FL, the participating parties may be curious to know each other's data, so a trusted third-party coordinator can protect the data confidentiality during the training process. However, if a certain level of trust exists between the participating parties, the need for a third party can be eliminated, and one of the parties can be the coordinator.

\vspace{0.2cm}
\subsubsection{\textbf{FTL}}
When the dataset of different clients slightly intersects in the feature and sample spaces, FTL (or hybrid learning) is the best candidate. FTL enables knowledge transfer from one domain to another, which helps in achieving better learning results. Specifically, a local model trained in one party is transferred to another party to leverage information extracted from the non-overlapping regions for enhanced model training at the other party. The most common example of transfer learning is the image classification problem. Several models exist that are tailored for classifying specific datasets, and they can be used to classify other types of datasets after making a minor tuning.

\subsection{FL Operation}
FL protocol consists of three main phases \cite{[5]} detailed as the following:

1. \textbf{Clients selection}: Albeit large-scale deployment is an attractive feature in FL, compared to classical ML, the number of clients participating in model training can easily reach thousands or millions of devices. This enormous number of endpoints reflects the capacity enhancements anticipated to be delivered by B5G (1 million/km\textsuperscript{2}) and 6G networks (100/m\textsuperscript{3}). As a result, end-device onboard capabilities and data distribution will vary considerably among the participants, rendering client selection a critical design aspect in FL. Several methods are proposed to address this issue, such as \cite{[3.12]}, where the authors propose a technique that improves the time-to-accuracy training performance by guiding the FL developers to select participants even at the scale of millions of clients. Further approaches are discussed in Section \ref{ClientsSelection}.

\begin{figure}
\centering
 \includegraphics[width=0.45\textwidth]{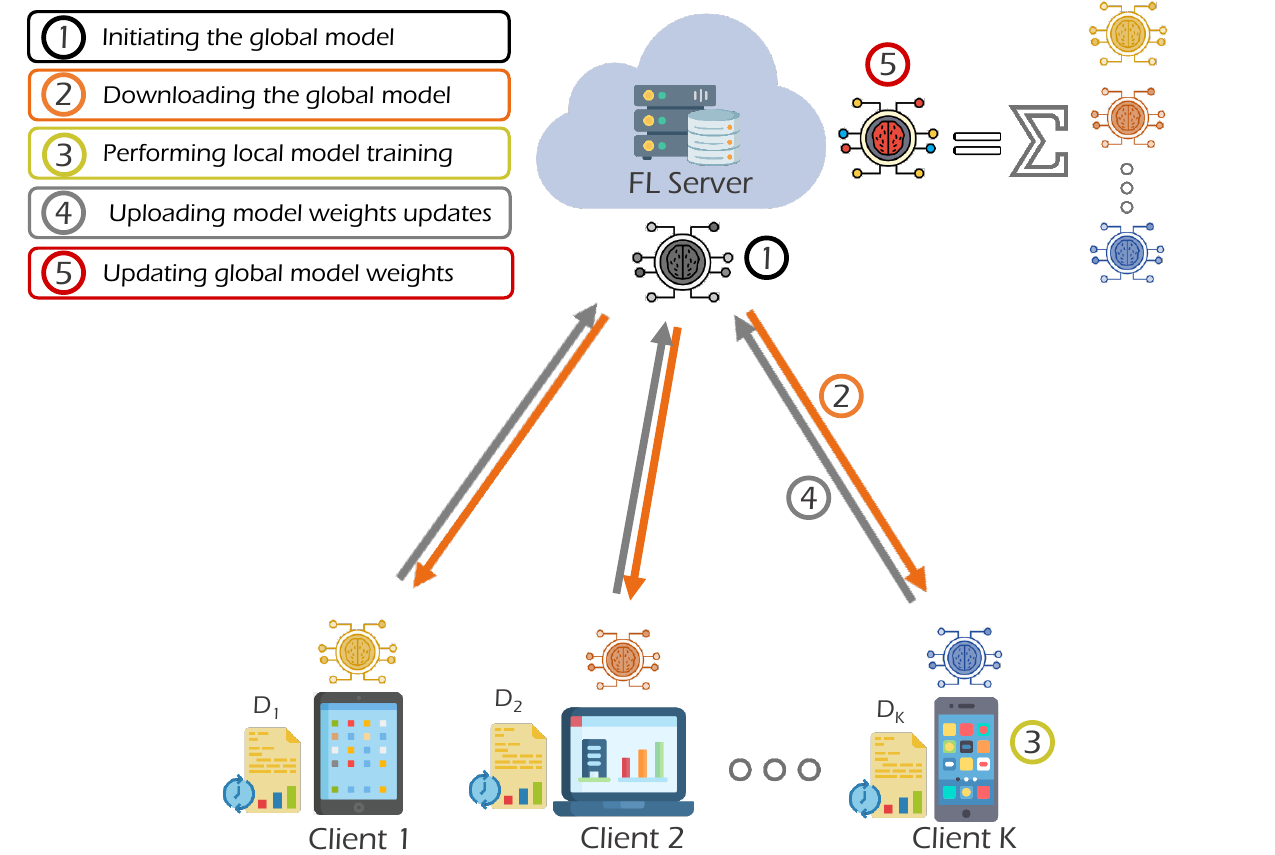}
 \caption{Sequential operation steps of FL involving $K$ participants.}
 \label{fl_operation}
\end{figure}

2. \textbf{Configuration}: In this phase, the selected participants receive the initial model parameters and train their local models based on the local datasets. In particular, after selecting participating devices successfully, $K$ edge nodes are ready to begin the training process. Fig. \ref{fl_operation} illustrates the FL's architecture and the operation steps. The device, $k\in\{1,2,...,K\}$, has a local dataset, $D_{k}\in\{D_{1},D_{2},...,D_{K}\}$, which includes input-output pairs of samples $\left(x_{i}, y_{i}\right), x_{i}, y_{i} \in \mathbb{R}$. In step {\large \textcircled{\small 1}}, the FL server initiates the global model created to perform a specific task and shares it with the selected participants. Next, at the $t$-th iteration, each participating node acquires the model weights $W_{t-1}$ and begins the model training by exploiting the data samples on their local storage. The objective of model training is to minimise the loss function $F_{k}(W_{t}^{k})$ of all data samples in the training dataset, $F_{k}(W_{t}^{k})=\frac{1}{D_{k}} \sum_{i \in D_{k}} f_{i}(W_{t}^{k})$, i.e., obtaining the optimum model parameters $W_{t}^{k}$ that minimise the loss function at each round of training which can be represented mathematically as, $\arg \min _{W_{t}^{k} \in \mathbb{R}} F_{k}(W_{t}^{k})$. Where $f_{i}(W_{t}^{k})$ indicates the loss on data sample $i$ given the parametrisation $W_{t}^{k}$, (steps {\large \textcircled{\small 2}},  {\large \textcircled{\small 3}}).

3. \textbf{Reporting}: At this point, the participants share the local model updates with the central server in a synchronous, or asynchronous manner \cite{[4.13]}. The synchronous strategy can produce a high-precision, fast-converging model in the absence of stragglers that are attributed to poor hardware or network resources. Stragglers threaten the scalability of FL as they slow down the training process. In contrast, the asynchronous mechanism handles stragglers naturally by incorporating participants' updates as soon as they arrive but has the threat of model quality degradation and insecure aggregation, resulting in an undesirable level of privacy. Several works have considered this issue by proposing different schemes, such as a secure buffer in FedBuff \cite{[12.1]}, staleness-awareness in FedSA \cite{[12.2]}, and semi-asynchronous in SAFA \cite{[12.3]}. Finally, the server aggregates the shared parameters to update the global model. Specifically, models aggregation and global model parameters computation are performed at the server as the following $W_{t}=\sum_{k=1}^{K} \frac{D_{k}}{D} W_{t}^{k}$, where D represents the entire dataset of all clients, i.e., $D=\sum_{k=1}^{K} D_{k}$, (steps {\large \textcircled{\small 4}},  {\large \textcircled{\small 5}}). The steps from {\large \textcircled{\small 2}} to {\large \textcircled{\small 5}} are repeated until the global model converges to a desired accuracy.

\subsection{FL Aggregation Schemes}
Gradient descent (GD) algorithm that aims to find the minimum of a differentiable function is commonly used in various ML algorithms, especially NN models \cite{[3.15]}. However, the computational complexity of GD increases with the dataset size, making it unsuitable for FL systems due to the slow convergence rate. An alternative to GD is stochastic GD (SGD), which can perform gradient calculation over a subset of data, significantly enhancing the convergence rate. In the FL setting, SGD (FedSGD) is exploited as an approach to quantify how often the global FL model needs to be updated \cite{[1.5]}. FedSGD is the basic aggregation scheme for FL-enabled systems, where clients compute gradients using random data samples. However, the FedSGD technique requires many communication rounds proportional to the volume of nodes’ datasets, which will burden the communication links and consume BW.

To address the above problem, the federated averaging (FedAvg) strategy has been proposed to alleviate the pressure on communication resources \cite{[1.5]}. FedAvg is a generalisation of FedSGD, where each node repeatedly runs SGD locally over different local data subsets and finds the optimum model parameters by averaging the locally evaluated gradients. Three main parameters control the performance of FedAvg: (i) the fraction of the selected nodes that perform computation at each round, (ii) the size of data subsets, and (iii) the number of epochs that the node passes over its dataset in every round. In FedAvg, instead of sending the computed gradients, each node will only send the model parameters. Thus, compared to FedSGD, the FedAvg algorithm performs more local computation and less communication with the server.

Nevertheless, in real-world scenarios, in which network devices are heterogeneous and the local datasets are non-identically distributed, FedAvg experiences poor convergence behaviour. Therefore, some variants of the FedAvg algorithm have been introduced to develop faster aggregation techniques. FedProx was proposed to solve the heterogeneity issue in federated networks \cite{[3.18]}. The FedProx principle is similar to that of FedAvg but with a small critical modification that improves performance. Instead of forcing every node to perform the same computation work, FedProx considers system heterogeneity by allowing each node to perform an amount of local computation proportional to its resources. Accordingly, enabling parameter aggregation from a set of heterogeneous nodes.

Another extension to the FedAvg scheme is the FedSplit algorithm \cite{[3.19]}, which relies on the operator splitting procedure for convex optimisation problems. Operator splitting is an efficient method for solving large-scale convex problems by performing iterations of simple and computationally inexpensive operations. It converts the problem into simpler sub-problems and makes progress on them separately. Motivated by the failure of FedAvg and FedProx to preserve the fixed points of the original optimisation problem, FedSplit is proposed as a splitting algorithm for federated optimisation to achieve rapid convergence. Moreover, the work in \cite{[3.21]} applied the adaptive optimisers ADAGRAD, ADAM, and YOGI in the FL setting, i.e., FedAdaGrad, FedAdam, and FedYogi. Extensive experimental evaluations are performed to examine these algorithms compared to the FedAvg algorithm. Furthermore, the Qsparse-local-SGD algorithm \cite{[3.20]} considers both local computation and communication reduction in distributed settings. Convergence analysis is made in synchronous and asynchronous FL, showing that the Qsparse-local-SGD algorithm achieves the same convergence rate as FedSGD. 

The above approaches are mainly designed for NN models where the parameters, i.e. weights and biases, are the main elements to update the global model. Despite numerous attempts to enhance the aggregation process, NN and DL models incur high communication and computation costs. Therefore, several studies have begun to explore other low-complexity techniques, such as ensemble learning under the FL settings, like FedBoost \cite{[11.3]} and FedTrees \cite{[11.4]}. It has been demonstrated in \cite{[11.3]} and \cite{[11.4]} that when the federated model is trained according to these algorithms, an excellent performance is achieved in terms of accuracy, computation time, and communication rounds. This paves the way for exploring other ML techniques in the FL environment.

\section{FL Enabling Technologies} \label{enabling tech.}
With the aim to realise the full potential of FL, several enabling technologies can be leveraged in order to improve the performance of FL and hence, accentuate its promising features. This section is devoted to discussing some of these enabling technologies.  

\subsection{Multi-Access Edge Computing (MEC)}
The rapid evolution of the internet-of-everything (IoE) paradigm has resulted in a plentitude of end devices. The abundance of resource-intensive devices, coupled with the emergence of QoE-oriented applications, has led to a wealth of data being generated at the edge of wireless networks. Exploiting this data requires sending it across the networks to reach the cloud server where the significant computation and storage resources are located. Accordingly, cloud computing has become unsuitable for resource-limited real-time applications, owing to the increased overhead occurring in the network, in terms of energy and spectrum resources, in addition to the increased latency and compromised security. Therefore, the European telecommunications standards institute (ETSI) has introduced a new computing paradigm called MEC, which brings cloud computing capabilities to the edge of the radio access network \cite{[3.22]}. The key motivation behind MEC is that running applications closer to the end-users with their associated computation tasks will reduce network congestion and preserve the network resources, enabling enhanced user experience.

The proliferation of smart end devices with the employment of MEC provides a suitable environment for employing FL algorithms. Shifting to decentralised ML model training at the network edge allows for greater scalability by distributing the computation from centralised architectures of the network core/cloud to the edge closer to the users. Moreover, MEC enables FL algorithms to offer latency optimisation for real-time applications where data aggregation, analytics, and computation are handled within user proximity. In fact, the capabilities of the edge server enable it to act as an FL server, whereas the widely dispersed edge devices are used as FL clients. Thus, MEC and FL provide rich services and applications close to the end users.

\subsection{Blockchain}
As a decentralised learning algorithm, FL has benefits in two main aspects: load balancing and privacy-preserving. However, FL has shortcomings as it does not keep records of participants' training contributions along with reliance on a central server prone to a single point of failure. In this regard, blockchain, a decentralised database managed by distributed nodes, can play an essential role in FL. Blockchain was initially introduced in 2009 as a type of distributed ledger technology \cite{[3.32]}. In particular, blockchain was primarily proposed to serve as a ledger of the public transactions for the cryptocurrency Bitcoin. For improved security, the principle operation of blockchain relies on grouping multiple transactions and storing them in a block encrypted by a hash signature. After that, each new block is time-stamped and chained with the previous one, creating a long chain of encrypted chronologically-ordered transactions. Therefore, blockchain has a high level of security, as altering the content of any block requires an agreement from all nodes connected to the chain. These merits motivate the authors in \cite{[5.2]} to design an incentive mechanism for a blockchain-enabled FL platform that can record and secure the workers' updates and reward them accordingly. Whereas the study in \cite{[7.1]} sheds light on the distributed ledger feature of blockchain to realise decentralised FL training without needing a central server, and proposes a new paradigm called FLchain.

Furthermore, the blockchain provides a fully transparent network in which all nodes can observe all transactions coming in and going out. When a new transaction is stored in the blockchain, it is considered immutable because it is verified based on a consensus mechanism. The consensus mechanism validates the data in each block and verifies its availability since all blocks will store the same copy of the data. This has been exploited in \cite{[5.3]} when the authors designed a blockchain-based FL system that can prevent malicious model updates using blockchain's immutability and decentralised trust property. Moreover, the work in \cite{[11.5]} uses blockchain to support the operation of the FL utilised to provide up-to-date service provisioning and support device communication in vehicular environments. Blockchain technology is adopted to ensure the credibility and integrity of sensitive services, such that the local models are verified using a consensus algorithm. Similarly, the authors in \cite{[47]} integrated blockchain with FL to maintain and secure local model parameters, improve learning quality, and optimise the allocation of varying resources in B5G networks. In light of the above discussion, decentralisation, availability, transparency, immutability, and security are the most promising features of the blockchain, which are well-suited for the FL system and constitute one of its enabling factors.

\subsection{Network Slicing (NS)}
NS is one of the key enablers in B5G wireless networks, where it exploits the network's physical structure to create several independent logical networks called slices \cite{[3.36]}. Each slice comprises an end-to-end isolated network tailored to fulfil diverse application requirements. In this respect, millimetre wave (mmWave) and terahertz communications (THz) in B5G/6G networks enable improved capacity for devices operating in a small coverage area, allowing the realisation of different IoE networks. These networks will require resources that meet the diverse quality-of-service (QoS) requirements. The unique characteristic of NS is that it grants each network segment an isolated and tailored slice to enable a particular service. However, configuration, activation, association, and management of network slices constitute a challenging factor that requires developing dedicated intelligent techniques. Therefore, using AI is a must to optimise real-time resource allocation and distribution among different slices according to their requirements. In light of this, FL and NS are considered promising enablers for each other. The work in \cite{[63]} presents an FL-based framework that predicts slices' service-oriented key performance indicators (KPIs). The concept of an in-slice manager was introduced for monitoring and collecting slices' KPIs and local decision-making to ensure optimal performance.

NS allows future mobile communications to ensure the efficient allocation of services while guaranteeing the QoS. For this reason, the study in \cite{[11.7]} proposes an FL-based forecasting algorithm to predict base station level traffic in sliced network architecture to facilitate intelligent and predictive management of resources. Whereas the authors in \cite{[11.8]} present a federated deep reinforcement learning  (DRL) scheme to manage the transmission power and spreading factor resources in LoRa-based industrial IoT (IIoT) slices. A multi-agent self-model is trained under the FL environment to obtain an optimal decision of LoRa parameters that fulfil the QoS of IIoT virtual network slices. Furthermore, the proposed work in \cite{[65]} offers a hybrid federated RL framework to find the optimal device association for radio access network (RAN) slices to maximise the network throughput.

\section{FL APPLICATIONS IN WIRELESS NETWORKS} \label{Apps}
\begin{figure*}[!t]
\centering
 \includegraphics[width=1\textwidth]{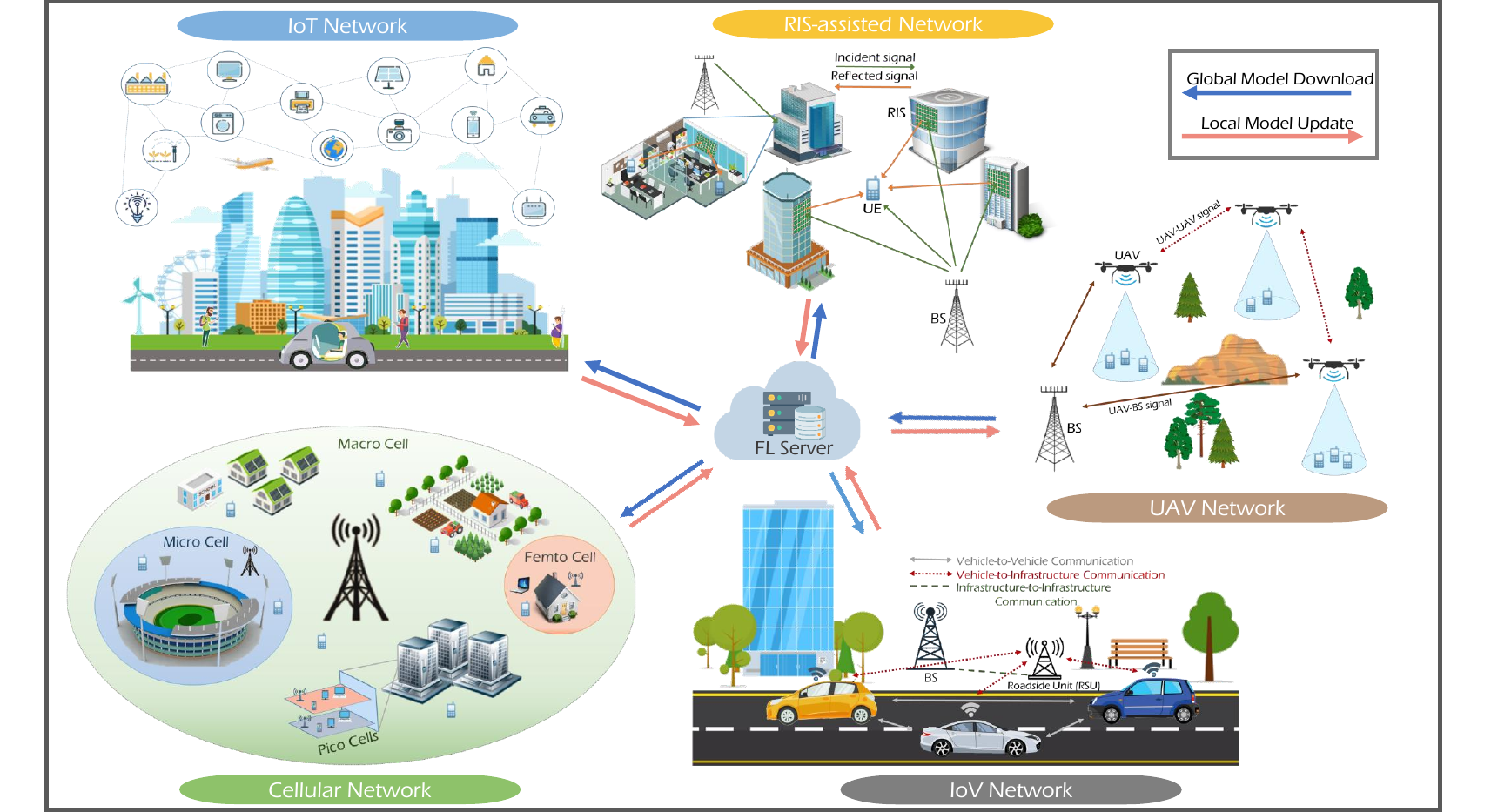}
 \caption{FL in various wireless networks; FL algorithm in the context of single or multiple wireless networks.}
 \label{FL_Apps}
\end{figure*}

Since the advent of FL by Google in 2016, extensive research has been conducted to promote, enhance, and determine the best usage of this decentralised learning algorithm. Wireless networks are one of the forerunners to adopt FL in their architecture, as depicted in Fig. \ref{FL_Apps}. This section will thoroughly present the key driving applications of FL in wireless networks; more specifically, Section \ref{Curr_apps} sheds light on the existing FL applications and accompanying challenges along with their potential solutions. Whereas Section \ref{potential} describes the significance of FL in new and promising application areas of the forthcoming B5G and 6G networks.

\subsection{FL State-of-the-Art Applications}\label{Curr_apps}
This section presents the research that has been done on utilising FL in current wireless networks.
\vspace{0.3cm}

\subsubsection{Cellular Networks}
\hfill
\vspace{0.1cm}

\indent The rollout of 5G in late 2020 has allowed operators to launch numerous commercial services that benefit from the enhanced features provided by this new technology \cite{[4.16]}. In addition, the use of FL in these networks has many applications in different areas, as described below.

\paragraph{\textbf{Homogeneous Cellular Networks}}
This type refers to low-frequency wireless networks with macrocells, alluding to their wide coverage. Two main concerns for FL at the network edge are heterogeneous devices with different computation and communication capabilities and securing local model updates. The work in \cite{[47]} presents a blockchain-enabled FL framework to ensure security in a trustless environment using a distributed ledger between entities. Blockchain is an intermediary between the FL server and edge nodes to verify model parameters based on the consensus process. Also, FL has applications for network function virtualisation (NFV), which is introduced as an innovative concept that enables adaptive resource allocation for future wireless networks. Subramanya \emph{et al.} \cite{[50]} leverages the FL technique to build a model that can proactively predict the auto-scaling setting for MEC virtual services and ensure data protection policies.

\paragraph{\textbf{Heterogeneous Cellular Networks}}
Heterogeneous networks (HetNets), which comprise different cell types, expand wireless networks' coverage and capacity. FL can be implemented in HetNets for resource allocation purposes. It was demonstrated in \cite{[48]} that applying HFL by grouping the users and assigning the needed resources for transmission can reduce the end-to-end communication latency in HetNets. This can be achieved by dividing users into clusters and assigning each cluster to the closest SBS. On the other hand, 5G HetNets are vulnerable to attacks, like denial of service (DoS), evil twinning, and port scanning. The work in \cite{[56]} proposes a node-edge-cloud framework empowered by HFL to detect attacks throughout the 5G HetNets. Specifically, multiple dedicated nodes are distributed inside the network, each of which performs model training by employing the RL technique to enable adaptive learning that can capture the rapidly changing nature of the HetNets environment. In addition, the work in \cite{[49]} presents FL-empowered MEC framework to tackle the communication overhead and delay between the edge server and clients in FL to enhance the training efficiency.

\paragraph{\textbf{Multiple-Input Multiple-Output (MIMO)}} FL has many applications associated with MIMO technology. Given the high dynamicity of mmWave systems, the study in \cite{[11.9]} enhances the performance of massive MIMO systems by estimating channel state information. FL is leveraged to conduct decentralised learning on the user side using local pilot signals to predict channel matrix, which helps determine the best beamforming design and improve the system’s performance. Moreover, the work in \cite{[11.10]} presents an energy-efficient solution to support multiple FL groups in future wireless systems. Massive MIMO is utilised to assist model updates and ensure a stable operation of multiple FL processes executed within the same coherence time.

\paragraph{\textbf{Fog-Cloud RAN}} The ever-increasing number of connected devices in 5G and beyond networks necessitates the transition to an ultra-efficient air interface. As a result, two air interface structures evolved, namely cloud-RAN (CRAN) and fog-RAN (FRAN). When surveying the literature, we observed many FL applications in FRAN networks, but using FL in CRAN networks is scarce. For example, the study in \cite{[11.11]} optimises the latency and BW resources when deploying FL in reconfigurable intelligent surface (RIS)-aided CRAN systems. The RIS controls channel propagation conditions and supports over-the-air computation (AirComp) technique to perform coherent on-air aggregation for local models by allowing simultaneous transmissions from clients to the parameter server. On the other hand, the FRAN paradigm fully uses edge networks and provides vital features such as content caching for optimal application performance and user experience. The authors in \cite{[11.12]} propose an FL-based mobility-aware content-caching framework in FRAN-based networks. Mobility and content demand statistics are exploited to improve users’ QoE by predicting and caching the most likely future content.

\paragraph{\textbf{5G New Radio (5G-NR)}} 5G-NR is a new radio interface standard designed by 3GPP to satisfy the growing demands of 5G mobile networks. This new radio access technology allows user equipment to switch dynamically between different resource blocks with different BWs. However, such a technique raises resource allocation challenges in B5G networks. FL has many applications in resource allocation in terms of computation, communication, and energy efficiency. For example, the study in \cite{[11.13]} uses FL to develop an ML model that aids in performing distributed resource management in cellular networks while minimising uplinks transmit power.

\subsubsection{Internet-of-Vehicular (IoV) Networks}~\\
\indent IoV has recently emerged as a key enabler for intelligent transportation systems (ITSs), combining two key concepts, namely, vehicle networking and intelligence \cite{[4.22]}. Within this context, the IoV paradigm aims to achieve smart information interaction between a vehicle and all network entities. Whereas vehicle computation capabilities realise vehicular intelligence by exploiting DL algorithms, cloud and edge computing, and big data analytics.

\vspace{0.1cm}
\paragraph{\textbf{FL in ITS}}
Communication reliability and latency are particularly significant in ITSs, owing to the severe consequences that might affect human safety. The proposed work in \cite{[34]} exploits the integration of FL with blockchain to realise a distributed, privacy-aware, and efficient model designed for autonomous vehicular networks. The diverse nature of the vehicles in ITSs is particularly appealing for FL applications. The heterogeneous data helps improve the model accuracy by incorporating all network scenarios experienced by different vehicles. In addition to latency, FL has shortcomings in server centralisation, where exchanging large updates between the participants and the server yield a high overhead on the server. To overcome this challenge, \cite{[34]} employs the blockchain technique, in which the distributed ledger is shared with each vehicle and maintains copies of the global and private models available and verified by each vehicle, relieving the pressure imposed on the central server. \cite{[9.3]} studies the use of FL setting within the context of URLLC in vehicular networks. It mainly focuses on proposing a distributed joint transmit power and resource allocation framework that can reduce the power consumption of vehicular users while ensuring low-latency communications. 

\vspace{0.1cm}
\paragraph{\textbf{Vehicular Edge Computing (VEC)}}
Following a similar concept to MEC, VEC exploits the communication and computation capabilities at the network edge. Ye \emph{et al.} \cite{[33]} implement FL with VEC to perform image classification to support diverse applications in ITSs. A model-selective approach was proposed to select clients with the highest computational capabilities and select models with the best image quality for aggregation. In an asymmetric FL setting, the server has no information about clients’ data and resources. To this end, a two-dimensional contract mechanism is proposed in \cite{[33]}, in which the server designs contract bundles that include various levels of data quality, computation capability, and rewards, and then the clients select the bundles that increase their utility. As part of IoV networks, electric vehicle (EV) networks are becoming more popular as the number of EVs increases; such networks are expected to take over from traditional vehicles in the coming years. The work in \cite{[36]} studies energy efficiency and profit maximisation at charging stations (CSs). It proposes an FL-based economically efficient framework to investigate the historical energy transactions to increase CSs profit. Specifically, FL is used to train a local model using CS private data to predict the EVs’ energy demands. After that, the local models of every CS are aggregated and shared amongst them to benefit from other CS information, yielding more accurate results.

\vspace{0.1cm}
\paragraph{\textbf{Traffic Perdiction}} 
Traffic prediction in smart cities brings up many benefits for ITSs, such as road safety, congestion avoidance, and shortest route selection. These gains are pronounced when exploiting information gathered from the edge in parallel with FL. One enhancement technique for FL is selecting the best hyperparameters of the local models in edge devices. As most of the literature focused on FL global optimisation, privacy, and communication, very few studied optimising model parameters. Qolomany \emph{et al.} \cite{[44]} proposed a particle swarm optimisation (PSO)-based technique to optimise local hyperparameters at the edge devices. Specifically, PSO optimises the local NN parameters, including the number of layers, neurons per layer, and epochs. This optimisation technique has been evaluated in traffic prediction as a use case. The work shows that the number of client-server communication rounds to find the best parameters is significantly reduced. This technique is attractive due to its low complexity implementation. However, its limitation lies in the reliance on a random search for the best initial parameters, which requires an unpredicted time that may affect the whole learning process.

\vspace{0.1cm}
\subsubsection{Unmanned Aerial Vehicle (UAV) Networks}~\\
\indent The flying vehicles in a UAV network have many attractive features, such as low cost, mobility flexibility, and ease of deployment, enabling them to participate in many tasks considered hard to perform. The application of AI algorithms and the recent advancements in UAV technology have widened the use-cases ambit of UAV networks \cite{[4.26]}.

\vspace{0.1cm}
\paragraph{\textbf{AI-empowered UAV}}
The interplay between AI and UAV networks opens a new horizon for exploiting UAVs in more complicated tasks; however, data security and privacy remain significant challenges. In UAV-enabled mobile crowdsensing (MCS) applications, FL is particularly appealing for preserving the privacy of sensed data. In this regard, the authors in \cite{[37]} integrated an FL-based UAV network with blockchain technology to eliminate the need for a central server. In addition, blockchain enhances FL network security by expulsing the adversary clients and sharing safe model updates between clients. On the other hand, the work in \cite{[38]} proposed an FL-enabled air quality monitoring framework for secure MCS. A UAV swarm is utilised to measure the air quality, and the sensed data is used to train a lightweight model to predict the air quality index. FL is considered a promising candidate that can exploit the data silos collected by different agencies to produce a global model while preserving data privacy.

Following the MEC concept, federated edge learning (FEEL) can potentially reduce the end-to-end latency and communication overhead in UAV networks. Yet, as demonstrated in \cite{[42]}, the efficient implementation of FEEL in UAV-based IoT networks is restrained by the battery lifetime of UAVs. In this respect, computation resource and BW allocation optimisation were formulated in \cite{[42]} to enhance the FEEL performance in a UAV network. Also, in \cite{[39]}, FL has been utilised as an aided technique to reduce the communication cost between multiple UAVs and a ground fusion centre in the context of image classification for remote area exploration missions.
\vspace{0.1cm}
\paragraph{\textbf{Flying Ad-hoc Networks (FANETs)}}
With the interest of accomplishing complicated tasks in UAV networks, UAVs are grouped in an Ad-hoc manner to create a local network, which allows UAVs to cooperate to perform joint tasks. Recent trajectory design and remote monitoring developments rely primarily on ML algorithms \cite{[4.29]}. To recall, such classical algorithms do not fit in the context of UAV networks due to their high mobility and constrained energy resources. FL was proposed to reduce the communication overhead as an efficient paradigm for FANETs, in which all participating UAVs collaborate to estimate the initial model parameters. Then, initial model parameters from all UAVs are shared and leveraged for local model training. A FEEL server is employed for model aggregation to exploit the local models to develop an enhanced global model. 

Attributed to the inherent non-centrality nature of FANETs, such networks are vulnerable to several security threats that intend to disrupt their functionality, such as impersonation and jamming attacks \cite{[4.30]}. Centralised attack detection and mitigation approaches are impractical, owing to the highly dynamic topology of FANETs. Thus, decentralised techniques are mandatory for such types of networks. To this end, in \cite{[40]}, an FL-based device jamming detection for UAVs in FANETs was proposed. In addition to the enhanced security, the framework in \cite{[40]} has considered the data heterogeneity issue between different UAVs. In particular, a Dempster-Shafer technique categorises UAV clients based on their data quality into groups. Then the FEEL server selects high-quality data group(s) for model training purposes.

\vspace{0.3cm}
\subsubsection{Reconfigurable Intelligent Surface (RIS)-Assisted Networks}~\\
\indent The emergence of numerous mmWave and THz applications has flagged several concerns attributed to the vulnerability of such applications to signal blockage and shadowing effects. Motivated by this and with the recent advancements in the solid-state industry, RISs have emerged as enablers of future wireless networks \cite{[4.31]}. An RIS, comprising several reflective elements, can be artificially engineered to control the electromagnetic properties of wireless signals and enable diverse functionalities, including wave splitting, reflection, absorption, etc. Leveraging an RIS is particularly beneficial in AirComp-enabled FL scenarios, in which some clients may be experiencing blockage or weak channel conditions, affecting the global model training quality \cite{[55],[80],[82]}. AirComp is a technique that exploits the superposition nature of the wireless channel to transmit simultaneous model updates from multiple clients. Section \ref{AirComp} covers the details of this technique. Yang \emph{et al.} \cite{[55]} use the AirComp technique assisted by RIS to boost fast global model aggregation, which reduces the required radio spectrum for parameter transmission since the clients collectively send their updates using the same channel. Also, to further enhance and boost the global model aggregation quality, an RIS is used to reduce aggregation errors by strengthening the quality of combined signals. In this respect, aiming to unleash the full potential of RIS in FL settings, Liu \emph{et al.} \cite{[80]} formulated a joint communication and learning optimisation problem by taking into consideration device selection, transceiver design, as well as RIS parameters. 

The aforementioned contributions have assumed perfect channel state information (CSI) at the server and clients’ sides. However, acquiring CSI at the transmitter (CSIT) is not always attainable due to dynamic channel conditions, leading to a significant delay in receiving the CSI information, thus curbing the FEEL global model convergence. The proposed work in \cite{[82]} investigated the CSIT-free over-the-air model aggregation based on RIS-assisted FEEL. The CSI at the transmitter side is assumed to be unavailable, while perfect CSI is assumed at the server side. Besides, the RIS adjusts and aligns the channel coefficients with the model aggregation weights. To this end, the successive channel coefficients are constrained, as a function of RIS phase shifts, to be proportional to the weights of the local models. Moreover, the received scaling factor is optimised by minimising the aggregation mean square error. To solve this optimisation problem, a difference-of-convex algorithm was adopted. Furthermore, RIS has proven its efficiency in converting wireless channels into a smart electromagnetic environment. To realise high-speed RIS-based communication, the authors in \cite{[81]} proposed two FL-based RIS optimisation schemes: RIS-assisted outdoor and indoor IoT mmWave communications. In the former scenario, the RIS controller is considered the FL server, while the user equipment (UE) is a client. The clients’ data represents the CSI corresponding to their location and optimum RIS configuration. The trained model is aimed to optimise the achievable rate to enable high-speed mmWave communications. The latter scenario considers an access point (AP) connected to multiple IoT devices assisted by RISs and acts as an FL server, while the RIS and IoT devices are considered clients. The FL model is trained based on location information and optimal RIS configuration. As a result, the trained FL model can achieve high transmission sum rates in IoT networks.

\vspace{0.3cm}
\subsubsection{IoT Networks}~\\
\indent High-dimensional data analytics will shift the traditional IoT paradigms from connected things to connected intelligence. It is envisaged that FL will be an indispensable tool in intelligent IoT-based applications, which are spreading in diverse fields \cite{[4.33],[4.34]}. In this section, we outline the usage of FL in various sectors associated with IoT networks.

\vspace{0.1cm}
\paragraph{\textbf { Industrial IoT (IIoT)}} The fourth industrial revolution (Industry 4.0) was triggered by the advancements in automation and manufacturing industries, coupled with the emergence of IIoT devices. Albeit the promising features of FL can be beneficial for IIoT networks, the upsurge number of nodes that may participate in the training process might produce colossal traffic that burdens the network. Reliable participant selection schemes can reduce network overhead and alleviate communication costs. The work in \cite{[52]} presents a budgeted client selection algorithm that enhances the global model accuracy by choosing the best clients. This algorithm finds $R$ clients with the best test accuracy based on the secretary problem. More specifically, clients are interviewed sequentially and marked as selected or rejected, and then these clients will be ranked from the best to the worst to facilitate the selection process. Another serious design aspect in FL-empowered IIoT networks is edge device failure, which causes severe fluctuations in production quality. The authors in \cite{[54]} shed light on such aspects and propose an anomaly detection framework that uses FL to train edge devices to predict abnormalities, enabling enhanced communication efficiency. 
\vspace{0.1cm}
\paragraph{\textbf { Healthcare applications}} FL has become very popular in the field of healthcare applications \cite{[4.40]}. Pandemics negatively affect human health and cause negative impacts on economics. Recently, Covid-19 swept the world, causing health problems and mortality. Covid-19's primary manifestation is pneumonia which is detected using X-ray scanning. ML can play a vital role in such medical cases, in which collected data can be exploited to train an ML model that can predict the infectious state. By emphasising that patient data across different medical centres should be handled privately, the FL setting is the natural option for such applications. Therefore, Liu \emph{et al.} \cite{[45]} applied FL to datasets of various clinical centres; FL clients exploited the available local X-ray images of Covid-19 cases at each hospital to train a model that helps practitioners to determine if a patient has been infected, without leaking any personal information.

\vspace{0.1cm}
\paragraph{\textbf{ Financial Perspective}} The financial sector plays a central role in all societies. In particular, the dependency on credit cards has exponentially increased in recent years, facilitating everyday life. Security attacks constitute a major threat to credit card systems, resulting in critical information leakage and money loss. Currently, banks utilise their datasets individually to develop centralised ML algorithms for fraud detection to mitigate such threats, but this was unavailing as the datasets did not help create an accurate model due to their insufficiency. To overcome this challenge, the work in \cite{[46]} presented a framework that depends on FL to build a fraud detection system that is collaboratively trained using datasets from multiple banks. The problem is that the number of fraudulent transactions is too small compared with legitimate transactions; this can obstruct FL performance. To this end, the synthetic minority over-sampling technique (SMOTE) is used to oversample the minority class by producing synthetic datasets that can be used to train the FL model for enhanced model inference.

\subsection{FL Potential Future Applications}\label{potential}
After describing FL and presenting its applications in various wireless networks, we outline some prospective application scenarios in new and promising areas. Our vision is primarily inspired by the applications anticipated to be inherent in B5G and 6G networks.

\vspace{0.3cm}
\subsubsection{Visible Light Communications (VLCs)}
\begin{figure}
\centering
 \includegraphics[width=0.5\textwidth]{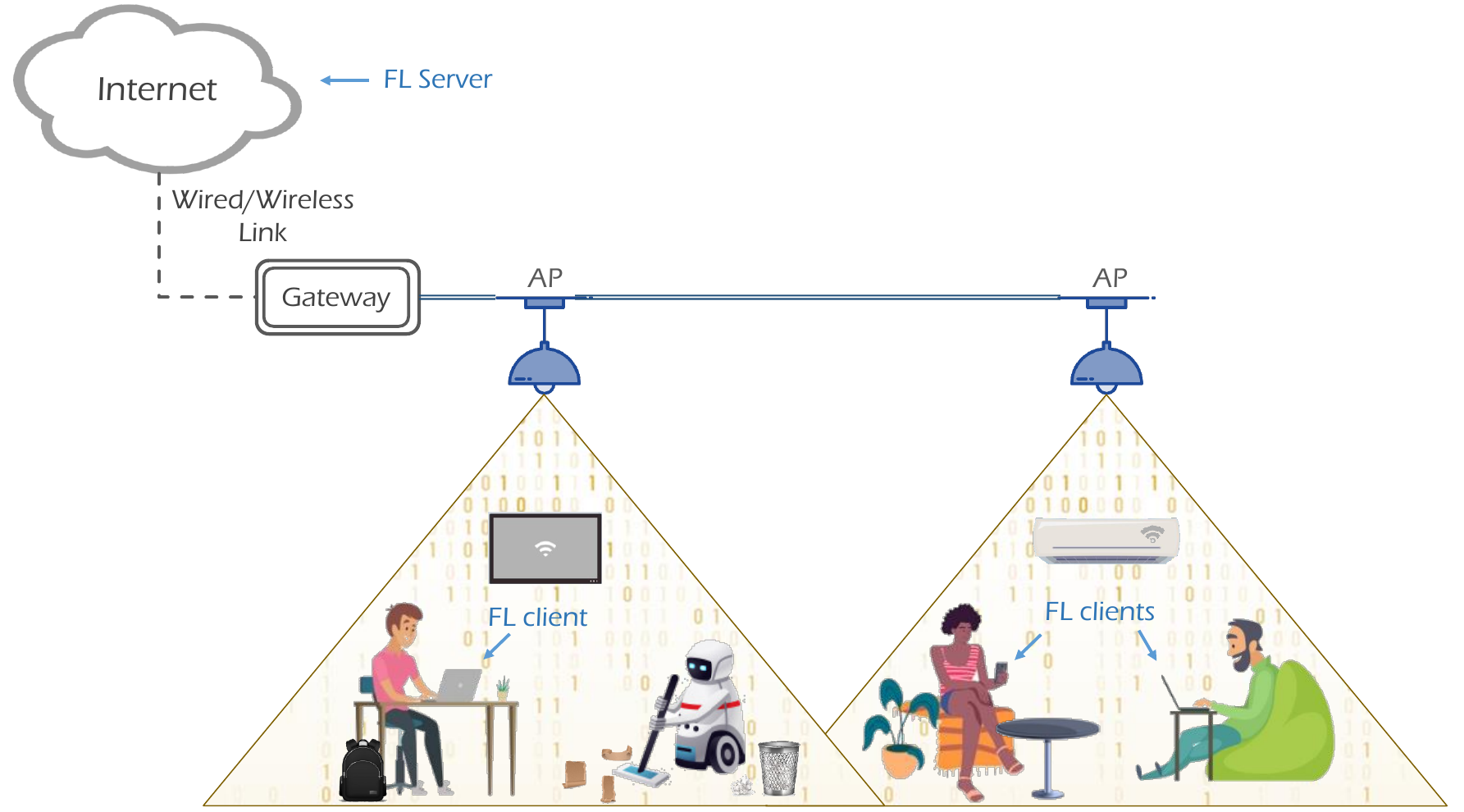}
 \caption{VLC system uses visible light as a medium for communication for wireless devices.}
 \label{VLC}
\end{figure}

~\\
\indent VLC is a new nascent wireless communication technology that relies on the visible spectrum for data transmission. VLC exploits the advantageous properties of light-emitting diodes (LEDs), such as low-power consumption, high brightness, and a long lifetime, to provide high data rate, low latency, and green indoor communications \cite{[3.50]}. VLC will play a major role in relieving the pressure on the scarce spectrum of the current wireless networks and provide a new connectivity method for the ever-increasing IoT devices. Fig. \ref{VLC} represents the VLC communication system that consists of LED units, called APs, connected to a gateway that, in turn, is connected to the external network through wired or wireless links. As a subfield of AI, FL will have a role to play in promoting VLC applications. The features that characterise VLC systems, including high spatial reuse, ultra-low-latency, ultra-high-data rates, and inherent security, provide the ingredients needed for the efficient implementation of FL algorithms \cite{[3.51]}. The main purpose of FL is to secure data privacy and reduce communication overhead. Accordingly, in the VLC network, the external server can be the FL server, and the deployed indoor devices can play the role of FL clients. In this case, clients may leverage the fast, secure, and reliable transmission environment to update the global model through the gateway, while the FL server can reach the required level of convergence faster.

FL training latency highly depends on client selection and scheduling. Therefore, how to properly select FL clients in the VLC network is an important question that needs to be addressed. Besides, as the number of participants increases, the global model can better infer accuracies. Nevertheless, the field of view of the LED units is limited and covers a limited number of clients so that a few devices can participate in the FL training process. To increase the number of participants in the VLC network, the HFL can be utilised, where the APs are used to aggregate the model updates of the clients under their coverage. Once this step is completed, the APs send the aggregated models to the central FL server. One interesting application of FL in VLC is predicting when an LED will stop illuminating due to, for example, LED life expiration or LED light-off time and instructing the endpoints to an alternate connection. Additionally, FL can play a major role in predicting clients' mobility, LED beam assignment, and client-LED association, to name a few.

\begin{figure}
\centering
 \includegraphics[width=0.423\textwidth]{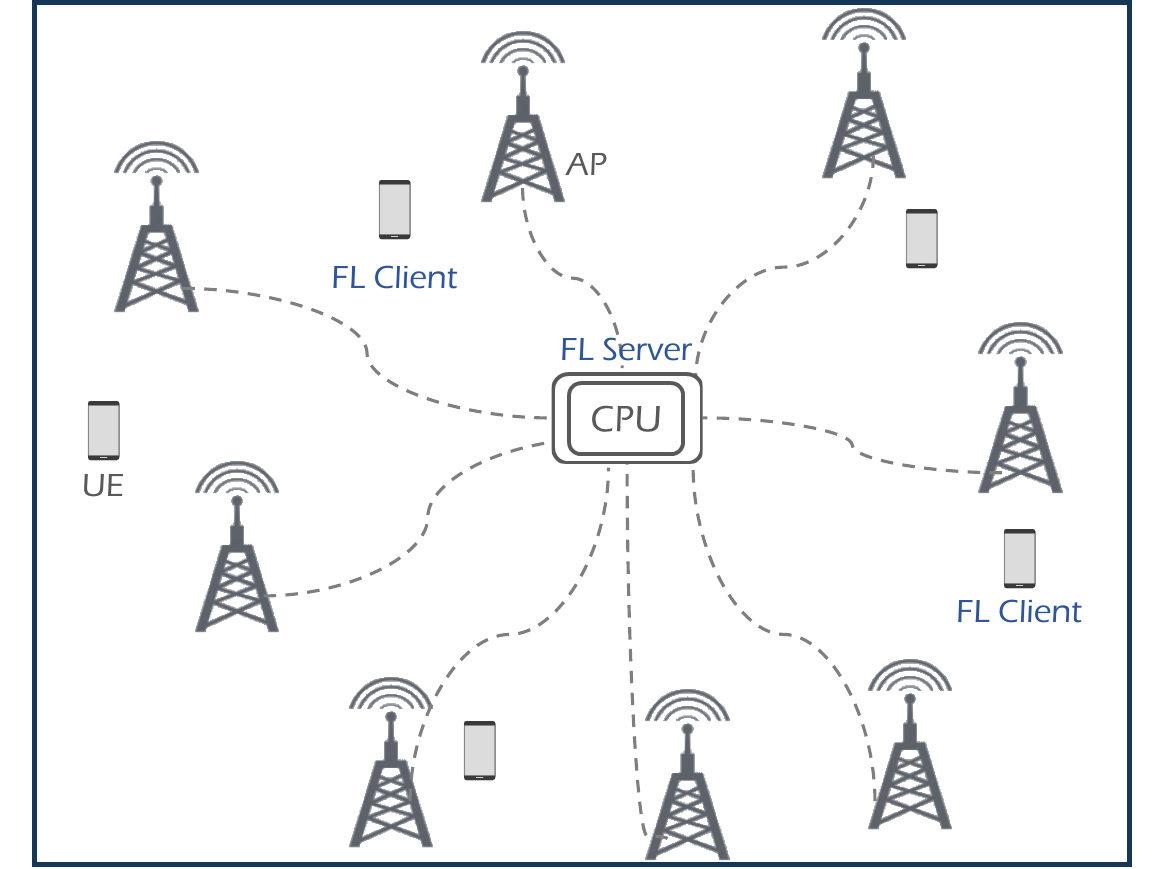}
 \caption{Cell-free mMIMO network which shows number of UEs surrounded by many APs.}
 \label{Cell-free}
\end{figure}

\vspace{0.3cm}
\subsubsection{Cell-Free Massive MIMO (CFmMIMO)}~\\
\indent The implementation of massive multiple-input multiple-output (mMIMO) networks includes two types based on the antenna deployment strategy: collocated and distributed antenna setup. The collocated type is easier to implement and has low data sharing overhead, which requires less backhaul. In contrast, the distributed implementation is more complex but gives the network an improved performance, especially in coverage gain. Recently, a new promising technology called CFmMIMO has been proposed as an incarnation of the distributed antenna setup \cite{[4.58]}. CFmMIMO constitutes a radical change in the cellular network paradigm as it eliminates the concept of cells. Many small, simple, geographically distributed BSs, called APs, jointly serve a small number of UEs using the same time-frequency resources via time division duplexing. APs are connected to a CPU through backhaul links and use the fronthaul to serve the UEs simultaneously, as shown in Fig. \ref{Cell-free}. CFmMIMO enhances the UEs connectivity by eliminating inter-cell interference and reducing the path attenuation due to the presence of the UEs near the APs.

CFmMIMO embraces distinct features that have a significant advantage in favour of FL. One such features is channel hardening \cite{[3.56]}, which means that the fading channel will behave as an almost deterministic scalar channel. Channel hardening greatly benefits FL, especially when selecting the clients to participate in the training process. Selecting UEs with stable connections eliminates any unfavourable transmission failure when uploading local updates to the FL server, i.e., CPU, thereby enhancing the FL performance. Moreover, when many APs surround the UEs, this will lead to high coverage gain and reduced distance between the UE and the AP. As a result, this will facilitate training the global model that requires a large number of clients to participate in the training process, thus reducing training latency and improving performance. Furthermore, FL can realise potential applications in CFmMIMO, for instance, creating FL models capable of assigning users to the optimal APs that fulfil the desired QoS by measuring the received signal strength of many surrounding APs. On the other hand, FL can be used to alleviate the congestion on the APs by training a model that can monitor, predict, and distribute UEs to APs in a way that maintains network performance.

\vspace{0.3cm}
\subsubsection{Satellite-Aerial-Terrestrial Networks}
\begin{figure}
\centering
 \includegraphics[width=0.5\textwidth]{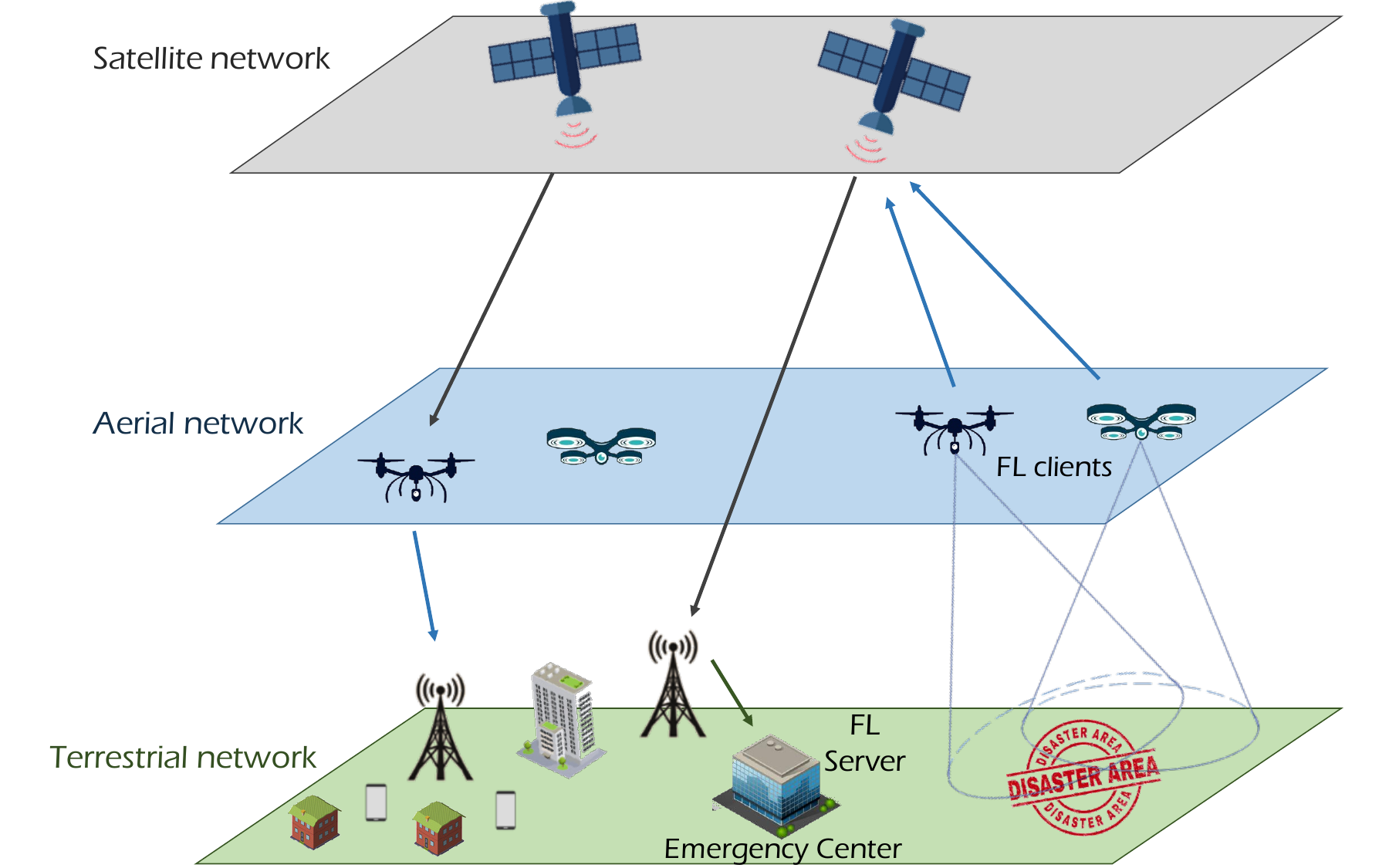}
 \caption{Illustration of satellite-aerial-terrestrial integrated networks. Satellites are used as a relay for communication between UAVs and terrestrial BS.}
 \label{Terrestrial}
\end{figure}

~\\
\indent Terrestrial cellular networks aim to serve populated regions while building such networks to serve sparsely populated areas like islands, oceans, and mountains is impractical. Satellite communication systems address this issue by providing rural areas with network connectivity. However, the quality of satellite links is not guaranteed due to challenges such as large path loss and limited UE power transmission. For this reason, the research has been directed toward utilising aerial platforms to aid satellite communications. High altitude platforms (HAPs) can be used to provide broadband services over a large coverage area \cite{[7.7]}. Moreover, HAPs provide more reliable communication links than terrestrial networks because they are less susceptible to ground blockages and multipath signal effects. Integrating the aerial-satellite network forms a space-backbone network layer that can provide wireless connectivity to ground users anywhere. As a result, the hybrid satellite-aerial-terrestrial networks have drawn the research community's attention for further improvements, which are envisioned to be an essential part of the B5G/6G networks. Fig. \ref{Terrestrial} represents the topology of the satellite-aerial-terrestrial network.

Recently, ML techniques have been considered in solving challenges related to satellite communications \cite{[7.2]}. Employing FL in satellite-aerial-terrestrial networks is still in its infancy; thus, there is plenty of room to explore the potential of FL in such networks. For instance, FL can tackle the network's limited resources, security, and energy usage challenges. Furthermore, satellite and aerial platforms have received significant attention due to their ability to deliver services in emergency scenarios such as disaster relief, and rescue missions \cite{[3.57]}. To achieve this, it is necessary to maintain robust and reliable communication between satellite, aerial, and ground-based networks. For instance, the terrestrial networks may be overloaded or destroyed if a large-scale disaster occurs, demanding a rapid establishment of a network to serve the afflicted area. Airborne vehicles can cover and monitor this area and send information to an emergency centre. However, in some cases, the vehicles may be outside the coverage of terrestrial BS; therefore, the vehicles can establish a connection with a satellite to act as a relay point between the air vehicle and the terrestrial BS, as demonstrated in Fig.\ref{Terrestrial}. Airborne vehicles provide the needed multimodal information; however, transferring a large amount of data burdens the communication links and consumes much time, which is critical in such situations. In this case, employing FL can eliminate the drawbacks. Equipping the vehicles with a pre-trained FL object detection and localisation model allows for sending lower-size vital information to locate survivors. Simultaneously, the vehicles can train the model using the collected data to enhance its accuracy and then send the model updates to the FL server. Accordingly, saving time and relieves communication links.

\subsubsection{Semantic Communication}~\\
\indent The main theme of communication systems up to 5G networks was to ensure the correct reception of every single transmitted bit, regardless of the meaning conveyed by the transmitted bits. However, this classical communication-theoretic framework does not meet the aspiration of B5G/6G networks, as the research community agrees on the need to upgrade this framework to a smarter and more informative one. The overlooked meaning behind transmitted data is expected to play a significant role in next-generation communication systems, forming an interface between machine intelligence and human intelligence. Therefore, considering data content's high-level meaning or relevance to support machine-intelligent services necessitates a shift from semantic-neural toward semantic communication systems \cite{[11.14]}.

The interplay between human beings and AI has resulted in many revolutionary applications like virtual reality (VR)/augmented reality (AR) and haptic communications. Several studies have begun to envision the integration of FL with semantic communications, VR/AR, and haptic communications. Similarly, in this article, we discuss several wireless scenarios where FL can be applied in these emerging fields. In semantic communications, using FL helps improve the network's BW utilisation by training a model that can extract relevant/contextual information from the data and filter out irrelevant information. FL-based semantic communication can effectively preserve the network's resources by transmitting semantic information rather than bits or symbols. On the other hand, VR/AR are real-time technologies that bridge the real and digital worlds by replacing or enhancing the physical environment with a computer-generated one. In the AR/VR environment, detecting users' movement and location is essential and heavily influences the wireless network's resources. FL is effective in predicting user movement and actions, which can be used to optimise the allocation of wireless resources to users \cite{[11.15]}. Finally, haptic communications bring a new dimension over conventional communication modalities by enabling real-time haptic experiences between tactile parties. Haptic communication will have diverse applications, particularly in industry and health sectors, which poses a critical need to protect such communications. FL is a vital tool for securing haptic-based applications through training an ML model that can discriminate between genuine and counterfeit actions based on previous signatures and warn the system of possible suspicious measures.

\section{FL CHALLENGES} \label{Challenges}
Deploying FL in various fields demonstrates its efficiency and highlights its main advantages. However, the successful implementation of FL is restricted by some challenges and limitations that must be resolved to realise its full potential. In this section, we articulate the most common challenges of FL and outline their proposed solutions. Table \ref{Comparison} summarises the key FL challenges and the associated solutions.

\subsection{Server Centralisation}
The performance of employing FL depends by large on the server and the participants. The bottleneck of either classical FL or HFL systems relies on the dependency on a centralised server to orchestrate the learning process, representing a single point of failure. Additionally, the large number of model updates sent to the central server can overwhelm the network, resulting in traffic congestion and degrading the network performance. Two approaches were used to address this challenge, namely blockchain and peer-to-peer. 

\vspace{0.1cm} 
\paragraph{\textbf{Blockchain approach}} 
Adopting FL systems integrated with blockchain instead of a central server avoids malfunctions that may result from using a single centralised server. Blockchain has been widely used in the literature \cite{[34],[37]}, where it can provide a distributed, end-to-end trustworthy training environment. The blockchain consists of miners and devices; miners can be randomly selected devices or separate nodes (such as cellular BSs or WiFi APs) that are computationally powerful to perform the mining process. The operation of blockchain-based FL systems can be summarised as follows: the process begins at the participating devices by computing and sending the local model updates to the associated miner in the blockchain network. Next, miners verify and exchange the local model updates using one of the consensus algorithms, generating a new block where the verified updates are recorded. Finally, the generated blocks that store the model updates are added to the blockchain and can be downloaded by the devices to perform the next round of computation. Leveraging the blockchain will not adversely impact the overall network system when a failure or malfunction happens in a miner, making the FL system more robust. 

\vspace{0.1cm}
\paragraph{\textbf{Peer-to-Peer approach}} The study in \cite{[72]} proposed a new technique called BrainTorrent, in which a centralised server is not required. This technique is aimed at medical applications where data sharing is prohibited due to privacy concerns. According to the authors in \cite{[72]}, BrainTorrent is a peer-to-peer procedure where each centre shares its model updates directly with the others without needing a central body to coordinate the process. Initially, every client maintains a version of the trained and old models. One of the clients in the network initiates the training process by sending ping requests to all other clients to update the model. Other clients will respond by sending their model weights and the training sample size. Then, the model weights are aggregated and averaged at the request initiator based on the clients' dataset size to produce a new version of the trained model, followed by repeating the process until a certain level of accuracy is attained. The main drawback of this technique is that it is feasible for networks containing a limited number of clients, while in an environment with a large set of clients, such a technique is impractical.

\begin{table*}[!t]
\centering
 \caption{Summary of FL challenges, impacts, and proposed solutions.}
 \includegraphics[width=0.88\textwidth]{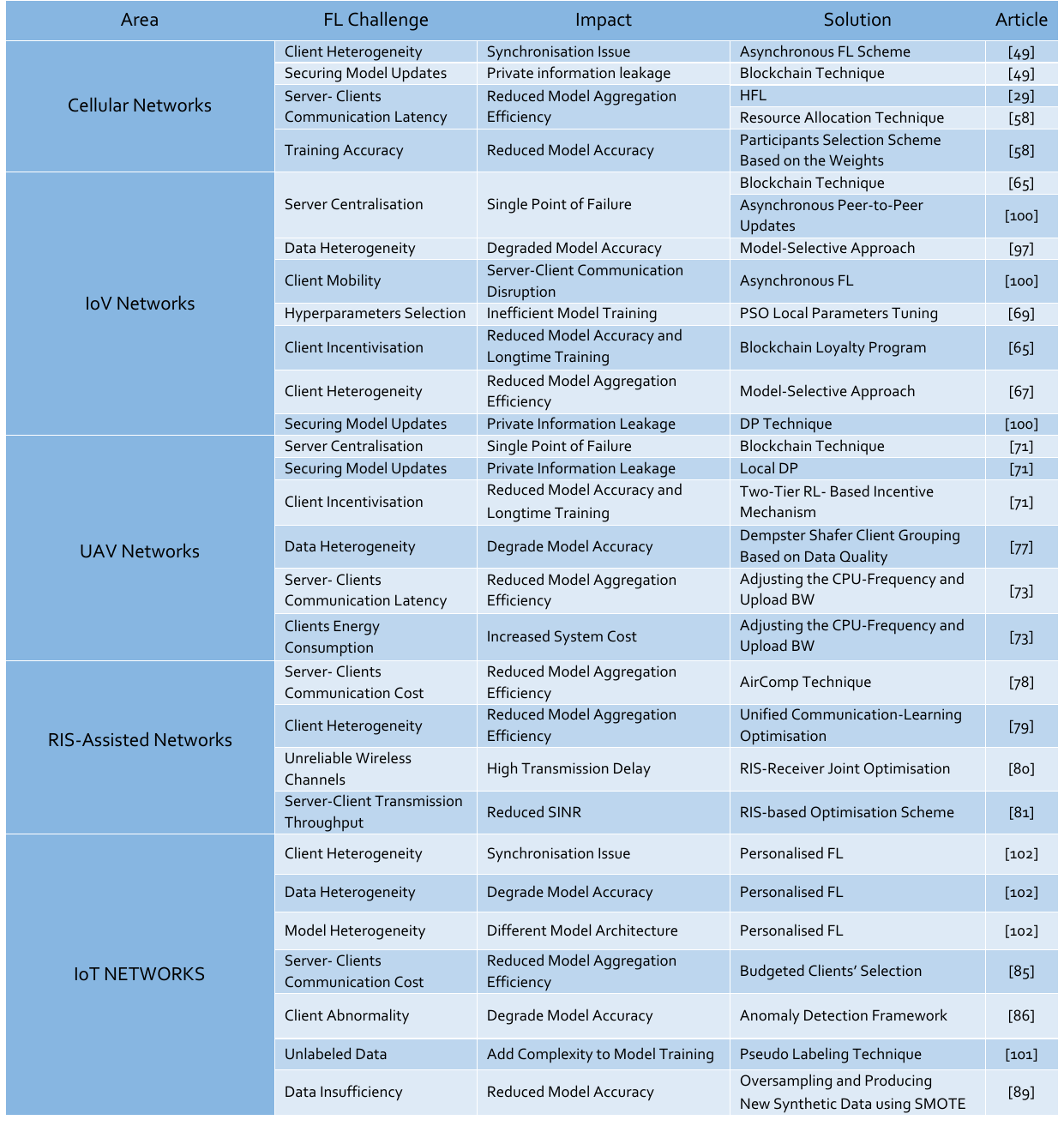}
 \label{Comparison}
\end{table*}

\subsection{Clients Selection}\label{ClientsSelection}
To recall, the FL process consists of three main phases: selection, configuration, and reporting \cite{[5]}. These three phases are performed iteratively until the FL model achieves a satisfying level of model accuracy. In the selection phase, the server determines the optimum users allowed to participate in the training process according to predefined selection criteria, i.e. whether or not the device is available and its resources. Concerning clients, two main factors directly impact the model convergence speed and efficiency.

\subsubsection{\textbf{Clients heterogeneity}}
In practical wireless networks, end devices have different hardware characteristics and experience varying channel transmission conditions, in addition to data heterogeneity. For instance, clients with high hardware capabilities and high-quality data can produce a well-trained model in a relatively short period compared to others. Furthermore, clients experiencing good transmission circumstances support low latency model transmission, enabling timely parameter aggregation. Failing to consider these aspects will reduce the efficiency of the FL training process. In \cite{[74]}, a participant selection scheme based on the available client resources has been proposed. Rather than selecting random clients, this scheme sends a resource request to the clients to collect information about their hardware specifications, communication reliability, and data availability. Based on this information, the server estimates the time required to complete a specific task and then selects clients that will participate in the following training round accordingly. In a similar context, the scheme in \cite{[33]} relies on selecting local models based on the clients' computation capability and data quality.
 
\subsubsection{\textbf{Clients incentivisation}}
FL depends on the participants and the on-device datasets. The high computation resources and valuable data attract FL to select these clients for training. However, nothing forces the end device to participate in a learning process that will deplete its resources and leads to unsolicited costs. Thus, a reward procedure must be considered to encourage the end devices to participate in the FL training process. In this context, various client incentivisation schemes are released \cite{[34],[37],[4.45]}. In \cite{[34]}, a loyalty program based on the blockchain technique is presented to motivate users with large samples of useful data to participate in FL training. According to their contribution, the loyalty program rewards the participants, attracting users with high data quality to participate in the training process. 
 
Furthermore, a two-tier RL-based incentive mechanism is presented in \cite{[37]}. The two-tier RL mechanism enables obtaining the best scenarios for the task publisher and clients in a dynamic environment by encouraging the workers to provide high-quality model training when explicit network parameters are unavailable. The reward of each client is maximised based on the contribution provided to enhance the global model. The work in \cite{[4.46]} improves the reliability of FL by proposing an incentivisation scheme that combines the client reputation and a contract theory to encourage clients with high-quality data to participate in model learning.

\subsection{Data Heterogeniety}
Data is the main driver of ML algorithms, and high-accuracy model training requires a large amount of data. Generally, practical datasets are heterogeneous and require pre-processing before they can be used for model training. 
\vspace{0.1cm}
\paragraph{\textbf{Data quality}} The characteristics of the locally generated data differ from one user to another. Datasets can be classified into two categories, independent and identically distributed (IID) and non-IID data. In practical scenarios, datasets are usually non-IID \cite{[4.47]}, while most of the existing literature in FL is based on the assumption of IID data. Given data heterogeneity, Ye \emph{et al.} \cite{[33]} employ a selective model aggregation approach to evaluate the quality of the images and then quantify it. The central server evaluates the image quality based on the clients’ historical records and prepares a contract to select fine clients with fine models. In \cite{[40]} Dempster\_Shafer technique is used at the global node to classify and prioritise the UAV clients into groups according to data quality. The highest priority group can contribute more to model training and produce better model weights. 

Fairness in FL has recently received more attention. As data heterogeneity increases among clients, the training process will produce a skewed model that may ignore some of the clients, resulting in fairness issues. A possible solution is to employ the personalisation concept \cite{[51]}, where the global model with coarse-grained features is sent to each participating device, and then the clients train the model using their data to build a model with fine-grained features. In \cite{[12.4]}, FairFed, a fairness-aware aggregation algorithm, is proposed. It relies on local debiasing techniques to give slightly higher weights to clients with similar local fairness to the global fairness metric, steering the next model update towards a fair global model. Likewise, the authors in \cite{[12.5]} consider fairness alongside the client selection problem and introduce a long-term fairness constraint to ensure that the average participation rate per client is no less than the expected guaranteed rate.
\vspace{0.1cm}
\paragraph{\textbf{Data insufficiency}} In some cases, the data collected by the devices may not be large enough to conduct model training. On the other hand, the percentage of high-quality data could be small compared to the total datasets, which affects the model inferencing and classification tasks. The proposed work in \cite{[46]} uses the SMOTE technique, which attempts to rebalance the classes in the datasets by oversampling the required features’ data. SMOTE generates new synthesised data examples close to the observed datasets. Another approach that can be used to provide more data samples is based on generative adversarial networks (GANs) \cite{[5.4]}. The goal of GAN models is to study and determine the distribution of the training data samples to generate more close to actual data samples from the estimated distribution.
\vspace{0.1cm}
\paragraph{\textbf{Data annotation}} Most studies considering FL assume a supervised training approach, where the data is processed and classified to facilitate the training process. However, in real situations, most of the generated data is unlabelled; in this case, the unsupervised FL is the method that should be considered. Data annotation is a challenging task that requires high cost and significant effort. The presented work in \cite{[43]} uses a pseudo-labelling technique to classify the unlabeled data based on the labelled data. Instead of manually labelling, which is time-consuming and requires much cost and effort, pseudo-labelling gives approximate labels depending on the model trained by the labelled data. The FL algorithm is used to train the model in two phases. First, the global model is trained by the distributed devices’ labelled data until reaching a certain convergence level. Second, improving the performance of the trained global model by training it again using the classified unlabeled data.

\subsection{Communication Cost}
The model convergence speed and accuracy in FL highly depend on the hardware specifications of the server and the clients. Despite the recent advancements in the computational and communication capabilities of end devices, model training and transmission overhead over multiple training rounds remain major design issue that potentially affects the global model training quality. Furthermore, a large number of model updates exchanged between the server and the clients can severely exhaust the network communication resources. In the following, we outline the main approaches to tackling this challenge.

\vspace{0.1cm}
\paragraph{\textbf{Models scaling and superposition}}\label{AirComp} Despite the significant advancements in edge computing, the lack of communication resources in current FL-enabled systems seriously affects the latency performance and reduces the model convergence rate. To this end, AirComp has been proposed to provide a co-design approach for the FL aggregation procedure by utilising the superposition nature of radio channels for simultaneously transmitting model updates from different clients \cite{[3.28]}. Therefore, improving communication efficiency by reducing the required BW resources and providing fast convergence. Later, a new variant of AirComp was introduced, called broadband analog aggregation (BAA) \cite{[3.29]} to cover wideband channels that can carry the multidimensional updates of local models. Furthermore, in \cite{[3.30]}, the authors propose a framework for model aggregation that relies on digital modulation. The proposed scheme utilises a single-bit gradient quantisation and quadrature amplitude modulation at the edge devices to achieve fast model convergence.

\vspace{0.1cm}
\paragraph{\textbf{Resource allocation}} It was demonstrated in \cite{[42]} that joint optimisation of onboard computation resources and BW allocation can be a promising solution to the computation/communication overhead in resource-constrained devices. This issue is more pronounced in ultra-dense networks, where an excessive number of model updates must be exchanged for global model convergence, and this further yields traffic congestion. Thus, selecting a subset of clients has proved its efficiency in tackling the communication cost problem. In particular, employing only clients with high-quality data in the training process can speed up the convergence rate. Subsequently, a reduced number of training rounds will be performed. Yao \emph{et al.}\cite{[4.55]} propose a two-stream model approach to reduce the FL communication cost. In this approach, the single model that was typically used to be trained by the clients is replaced by a two-stream model. The authors exploit the transfer learning mechanism and maximum mean discrepancy to force nodes to learn other nodes' knowledge. The experimental results showed a reduction in the required communication rounds and reduced communication costs.

\vspace{0.1cm}
\paragraph{\textbf{Gradients compression}} Large-scale deployment of FL requires significant communication rounds between the central server and the clients. This necessitates expensive network resources to perform model parameter exchange, which can limit the scalability of the FL system. The work in \cite{[73]} significantly reduces communication costs by using deep gradient compression. The work shows that most SGD parameters are redundant, so the compression technique is employed to considerably reduce the number of transmitted parameters while preserving the model's accuracy. Moreover, the compression minimises the gradients by sending only the necessary gradients to the central server in each round, reducing communication latency and alleviating the utilisation of limited wireless resources. On the other hand, by exploiting non-orthogonal multiple access (NOMA), a new 5G medium access technology that improves spectrum efficiency by allowing simultaneous transmission over the same channel, the work in \cite{[4.52]} proposes a NOMA-enabled adaptive gradient compression FL system. In this work, the authors exploit NOMA and adaptive gradient quantisation and sparsification to facilitate uploading model updates over fading wireless channels.

\subsection{FL Latency and Convergence} \label{FL lat.& Conv.}
The network and devices heterogeneity, data statistics heterogeneity, dynamic wireless environment, and acquiring the CSI are the most important factors influencing FL performance in terms of latency and convergence rate. An appropriate client scheduling mechanism can be the key to an accurate and fast model convergence. The authors in \cite{[3.39]} formulate an optimisation problem that jointly selects a group of clients with local models that significantly impact the global model and assigns the limited resource blocks to those clients. Furthermore, Huang \emph{et al.} \cite{[3.40]} proposed a stochastic client selection algorithm that jointly considers the cumulative effect of participants and selection fairness to maintain a high-quality training performance while ensuring fairness among high-qualified and low-qualified clients. Moreover, enabling edge computing can remarkably reduce the FL latency, in which APs are placed close to the edge device, and hence, reduced latency can be achieved. Within the same context, the authors in \cite{[3.38]} proposed a framework to reduce the average time per round by considering latency-based scheduling, in which clients are selected based on their computation and communication delay.

Generally, ML algorithms are sensitive to hyperparameters, which play a critical role in the model convergence rate. Therefore, to further reduce FL training latency and enhance the convergence time, careful consideration should be taken in the design of efficient hyperparameters. In this regard, several hyperparameter tuning algorithms have been developed to manage many of these parameters with their wide ranges. This includes Bayesian optimisation, grid search, and random search. The work in \cite{[3.47]} develops a scheme that can efficiently determine the optimum learning rate (LR) values. In the proposed technique, referred as cyclical LR (CLR), the CLR is bounded by a range of carefully selected values, in which its value can vary. This approach aims to avoid random LR initialisation. The presented results in \cite{[3.47]} showed the efficiency of such a technique in reducing the FL latency by minimising the number of training operations while ensuring a particular level of accuracy.

\subsection{Securing Model Updates}
Although FL is motivated by inherent privacy-preserving and security features, sophisticated intruders can retrieve critical information about the participating nodes from the shared model updates. Besides, malicious devices may opt to participate in the training model process to inject false model updates, affecting the accuracy of the trained model. The following approaches are developed in order to ensure secure model transmission:

\vspace{0.1cm}
\subsubsection{\textbf{Secure multi-party computation (SMC)}} A cryptographic protocol that aims to conceal personal information and guarantee zero-knowledge between multiple involved parties \cite{[9.16]}. Its main idea is to distribute the computation between multiple parties without exposing or moving private information. Its working mechanism can be summarised as follows: first, the participated organisations' datasets are split and masked by adding random numbers, and then these encoded segments are shared between organisations to perform the required computation, thus guaranteeing data privacy and trust. SMC allows organisations to work together without knowing one anothers' confidential information.

\vspace{0.1cm}
\subsubsection{\textbf{Differential privacy (DP)}} This approach prevents leaking model parameters to intruders by leveraging artificial noise, which is added to the locally trained model before transmission \cite{[9.18]}. However, enhanced security comes at the expense of model accuracy; hence, joint optimisation is essential to strike a balance between security and model accuracy. Such technique has been used in the literature, e.g., \cite{[35]} and \cite{[37]}, in which random Gaussian noise is utilised to enhance the privacy of model parameters.

\vspace{0.1cm}
\subsubsection{\textbf{Homomorphic encryption (HE)}} HE is a key-based security mechanism which allows performing calculations on encrypted data. In the context of FL, participating clients generate public and private keys, where the former is used to encrypt locally trained models. After that, the model updates received from all clients are aggregated on the server side in an encrypted mode. The clients leverage the private keys in order to decrypt the global model updates. Albeit the enhanced security achieved by exploiting the HE mechanism, the computation complexity of the cryptographic operations imposes additional overhead on the resource-constrained clients in terms of time, power consumption, and communication cost. In this regard, Zhang \emph{et al.} \cite{[76]} proposed a batch encryption
technique, which minimises the encryption and communication cost resulting when using HE. Specifically, each client quantises the gradients to be represented in a low-bit integer format, and then a batch of the encoded gradients is encrypted for transmission. Consequently, the encryption overhead and the size of the total ciphertext will be considerably decreased.

\section{FUTURE RESEARCH DIRECTIONS} \label{Discussions}
Despite the prospects brought by the advancements of FL, its application is still in its early stages. This necessitates dedicating the research efforts toward addressing the associated challenges and exploring new horizons of implementation possibilities. In the following, we list a number of interesting future research directions.

\subsection{Data Freshness} In information technology, data is marked by the date of its creation and can become meaningless, i.e., outdated. Access to timely information (i.e., data freshness) is paramount for time-based systems driven by datasets \cite{[4.62]}. In order to quantify the data freshness, the age of information (AoI) metric is introduced \cite{[4.64]}, and is considered an essential parameter in realistic scenarios of data networks. From the perspective of FL, AoI can be defined as the time that elapses between collecting data from clients and completing the FL training task. Considering applications with tight latency and throughput requirements, e.g., ITS, the AoI becomes crucial in network design principles. Accordingly, future research may focus on proposing novel schemes that select FL clients based on their data freshness to ensure that the required network reliability is achieved. Additionally, distributed client datasets can be highly temporal and change rapidly; thus, incorporating the rapidly changing data and determining the correct timing of model updates is essential to enhance FL performance in highly dynamic environments.

\subsection{Spectrum Sharing} The widespread use of IoT devices and the new technological trends make the limited spectrum bands insufficient to meet the requirements of BW-hungry applications. To this end, spectrum sharing is proposed to mitigate the pressure on frequency bands by allowing multiple networks to operate using the same portions of the licenced or unlicensed spectrum, provided that they do not interfere with each other \cite{[4.67]}. Coexisting networks should consider interference problems, i.e., co-channel and adjacent channel interference, addressed by imposing strict rules from telecom regulators. Multiple networks from the same or different technologies can coexist and use the same spectrum band, where this coexistence is categorised into equal and different access rights. The major concerns associated with equal rights coexisted networks are maintaining seamless operation, mitigating harmful interference between them, and ensuring fairness. By exploring the literature, we conclude that it is difficult to satisfy these concerns without the intervention of a third party who must receive information from the coexisting networks and manage transmissions. However, this method is undesirable as it requires information disclosure and incurs additional communication costs. Therefore, the FL algorithm is a potential solution that preserves network data privacy and eliminates the need for a third party. Coexisted networks transmission demands and local spectrum utilisation can collaboratively train a global FL model, for instance, deep RL, to address coexisting issues. This model is fed back to each network to make the right spectrum access decisions.

\subsection{FL at Scale} The applications mentioned in Section \ref{Curr_apps} are considered small-scale scenarios. However, many applications require a wide deployment of FL to take advantage of the data collected in different locations. This helps to get feature-rich datasets from extensive scenarios that can train an effective global model. Designing an FL system that covers large-scale environments requires special attention to the FL server capabilities in addition to cellular and backhaul communications. The number of participants can easily reach millions spread in broad areas and produce massive model updates that must be transmitted through the wireless network. Therefore, considering the network's communication efficiency alongside selecting an FL server with efficient hardware to handle enormous amounts of updates is crucial. To this end, future research should consider wireless network design and the specifications of the FL server suitable for large-scale deployments and develop a technique that intelligently selects the optimum participants among many devices willing to participate promptly. 

\subsection{Meta-Learning} The shortcoming of existing ML techniques, especially DL algorithms, is that they rely on large datasets to develop a good model. In most cases, it is not possible to obtain a high amount of dataset, while in other cases, the number of samples that hold the desired features is small compared to the entire dataset. Therefore, finding a mechanism to train models based only on a small dataset sample is necessary. In light of the preceding discussion, the meta-learning technique is introduced to address data insufficiency \cite{[4.72]}. Meta-learning, also known as learn to learn, uses the metadata of other tasks, like data patterns, properties of the learning problem, and the algorithm performance to learn how to learn and then learn the new task more efficiently from a small set of data. This new learning method in the FL algorithm is expected to improve its performance in several aspects. First, using only a small amount of data samples in meta-learning leads to a more convenient client selection. Moreover, the operating cost will be reduced, thus saving many resources and training time. The optimal client selection will also lead to rapid model convergence and lower latency which is crucial for B5G and 6G networks. Finally, meta-learning can help adapt the global model to each user, especially when data heterogeneity exists among clients.

\subsection{Modality Agnostic Learning} In current ML approaches, the models are designed based on the characteristics of input data dedicated to a specific task. However, B5G/6G networks allow the creation of different dataset modalities, such as vision, audio, time series, and point cloud. When a specific model needs to be used with a different data configuration, its architecture must be redesigned. This means that best-practice models cannot be used in different domains without modification. Perceiver \cite{[5.6]} is an interesting solution proposed to handle the configuration of different data shapes based on Transformers networks \cite{[5.7]}, which are sequence transduction models that rely entirely on the attention mechanism. The usage of Transformers in computer vision has shown their efficiency in classification tasks using considerably lower computation resources. Therefore, utilising dynamic models that suit multimodal inputs, like Perceiver, in the FL setting will help in its realisation and wide adoption to perform different network optimisation tasks that render the network more reliable. In more detail, dynamic models will allow clients with different data shapes to participate in the FL process, facilitating FL operation. This new research direction needs further investigation under the umbrella of FL systems.

\subsection{FL Carbon Footprint} DL-based approaches are highly dependent on heavy computations, resulting in high power consumption. Higher energy cost increases carbon dioxide equivalent (CO$_2$e) emissions, constituting the main reason for climate change \cite{[9.20]}. Recent studies have been devoted to investigating the impact of ML on Earth’s climate, steering the focus to the environmental effects of training large-scale ML models connected to network grids powered using fossil fuels. The environmental consequences of FL in wireless networks have not been explored much; few studies have recently begun to detail such implications. In addition, the transition from centralised to distributed learning seems more energy efficient. Avoiding transmitting big data to a central location saves much network energy and eliminates the need for cooling and other auxiliary tasks. However, the ML technique and the number of participants determine how efficient the network is. With this in mind, the study in \cite{[9.29]} proposes a sustainable FL-based framework by considering energy harvesting technology. Our vision is that future wireless networks will highly depend on renewable energy resources; for instance, we may see more dependence on solar power at the edge devices. This aspect opens the horizons for exploring FL approaches that can potentially contribute to achieving carbon-friendly wireless networks. To this end, future research should be dedicated to assessing the environmental impacts of FL-empowered networks before being widely used in broader scopes.

\subsection{Low-Precision FL} Computational capabilities are a significant factor in determining the best clients involved in the FL process. However, edge devices often have limited computing resources, making implementing FL more complex. For instance, the computational complexity of DL models increases as the model becomes deeper, requiring high-performance hardware, while in reality, resource-limited devices are available. The use of full-precision DL models that perform floating-point mathematical operations is a major reason for the increased computational complexity of such models. Various approaches are introduced to compress deep networks, such as parameter pruning \cite{[9.7]} and parameter quantisation \cite{[9.10]}. Much interest has focused on the model quantisation technique as it produces more compact models than their floating-point counterparts. Binary neural networks (BNN) \cite{[9.13]} is a promising approach that recently emerged to facilitate deploying DL models in resource-limited devices. In BNNs, model weights are quantised using binary values. The merits of BNNs represented in memory saving, computation reduction, and energy efficiency make them appealing for use under the FL setting. The combination of FL and BNNs will form a new low-precision framework that can be used at the edge of wireless networks. Although the usage of BNNs addresses the scalability of the FL process, their performance is degraded compared to other full-precision counterparts. Using BNNs in FL is a promising solution; nonetheless, more research should focus on optimising BNN-based FL frameworks and closing the performance gap.

\subsection{Digital Twin (DT)} DT is a technology representing a physical object, service, or even an entire system in its counterpart digital version \cite{[7.8]}. The DT framework aids the operation of complex systems by providing insights into how these assets behave under various simulated circumstances that will help improve decision-making and optimise these systems. As reported by Gartner, DT is envisioned to be one of the most influential industry 4.0 technologies in the next decade. Furthermore, DT is a data-driven technology that can provide system operation excellency by leveraging real-time analysis when paired with AI. However, The DT faces the challenges associated with big data and privacy protection. Accordingly, a novel collaborative paradigm can be achieved when fusing FL with DT systems to meet these challenges. As two emerging and promising techniques, FL and DT can help reduce wireless networks' operation complexity and realise 6G-based IoE applications \cite{[10.4]}. Despite the literature's scarcity of works that leverage such fusion, it is envisioned to become an essential part of the next generations of wireless networks. Future research may consider using FL with DT to share knowledge between DT nodes and develop a common understanding. In addition, the DT can assist FL tasks, for example, by quantifying the DT node's trust and selecting clients based on the degree of trust. 

\subsection{FL Task-Reward Announcement} FL model training depends on the participating clients’ resources and the corresponding on-device datasets. Data quality differs from client to client based on usage, and behaviour \cite{[4.81]}. The selection of devices is based on predefined conditions like being connected to an unmetered network, idle, and in a charging state. Moreover, choosing the optimal clients for a particular task helps relieve the pressure on wireless spectrum resources by lowering the training rounds required in FL and improving network latency. However, to encourage users to participate in the FL process, a reward mechanism should be developed to compensate for their consumed resources and data used while training. How to determine and select participants based on their resources and data quality is ongoing research. FL task-reward announcement is a necessary approach. With an effective announcement technique, users with high-quality data and resources may be encouraged to make themselves ready to participate in the FL process by matching the terms of participation required to receive some rewards. Announcement techniques can enhance the overall performance of the FL system.

\section{CONCLUSIONS} \label{Conclusion}
The emergence of FL and its distinctive features pave the way for numerous advancements in the industry. Motivated by the various implementation scenarios in different wireless networks, we conducted a survey demonstrating the salient merits of FL. In this context, this review paper presented the basic operational principles of FL and discussed the essential enabling technologies. This is followed by a discussion of state-of-the-art wireless network applications optimised by utilising the FL mechanism. Moreover, we shed light on promising research directions that may unlock the potential of FL in new areas of B5G and 6G wireless communication systems. Furthermore, we focused on the challenges associated with implementing FL and outlined the techniques used to address those challenges in literature, and then we offered insights to improve the design of the FL algorithm. We believe that the way this survey is harmonised can offer a firm understanding of FL usage in various areas, facilitating the focus on new research directions.

\section*{Acknowledgement}
This article is supported by Ajman University Internal Research Grant No. 2022-IRG-ENIT-18. The research findings presented in this article are solely the author(s) responsibility.

\balance
\bibliographystyle{IEEEtran}
%\bstctlcite{BSTcontrol}
\bibliography{Survey}

\begin{thebibliography}{100}
\providecommand{\url}[1]{#1}
\csname url@samestyle\endcsname
\providecommand{\newblock}{\relax}
\providecommand{\bibinfo}[2]{#2}
\providecommand{\BIBentrySTDinterwordspacing}{\spaceskip=0pt\relax}
\providecommand{\BIBentryALTinterwordstretchfactor}{4}
\providecommand{\BIBentryALTinterwordspacing}{\spaceskip=\fontdimen2\font plus
\BIBentryALTinterwordstretchfactor\fontdimen3\font minus
  \fontdimen4\font\relax}
\providecommand{\BIBforeignlanguage}[2]{{%
\expandafter\ifx\csname l@#1\endcsname\relax
\typeout{** WARNING: IEEEtran.bst: No hyphenation pattern has been}%
\typeout{** loaded for the language `#1'. Using the pattern for}%
\typeout{** the default language instead.}%
\else
\language=\csname l@#1\endcsname
\fi
#2}}
\providecommand{\BIBdecl}{\relax}
\BIBdecl

\bibitem{[3.63]}
D.~Gil \emph{et~al.}, ``Internet of things: A review of surveys based on
  context aware intelligent services,'' \emph{Sensors}, vol.~16, no.~7, p.
  1069, July 2016.

\bibitem{[9.30]}
Z.~Rehena, ``{Internet of Things},'' \emph{Interoperability IoT Smart Syst.},
  p.~1, Dec. 2020.

\bibitem{[1.2]}
M.~Shafi \emph{et~al.}, ``{5G}: A tutorial overview of standards, trials,
  challenges, deployment, and practice,'' \emph{IEEE J. Sel. Areas Commun.},
  vol.~25, no.~6, pp. 1201--1221, Apr. 2017.

\bibitem{[1.3]}
H.~N. Dai \emph{et~al.}, ``Big data analytics for large scale wireless
  networks: Challenges and opportunities,'' \emph{ACM Comput. Surv.}, vol.~52,
  no.~5, pp. 1--36, Sept. 2019.

\bibitem{[3.67]}
M.~Obschonka and D.~B. Audretsch, ``Artificial intelligence and big data in
  entrepreneurship: a new era has begun,'' \emph{Small Business Economics}, pp.
  1--11, June 2019.

\bibitem{[11.6]}
P.~P. Shinde and S.~Shah, ``A review of machine learning and deep learning
  applications,'' in \emph{Proc. Fourth int. conf. comput. commun. control
  automat. (ICCUBEA), Pune, India}, Aug. 2018, pp. 1--6.

\bibitem{[4.5]}
P.~Li \emph{et~al.}, ``Multi-key privacy-preserving deep learning in cloud
  computing,'' \emph{Future Generation Comput. Syst.}, vol.~74, pp. 76--85,
  Sept. 2017.

\bibitem{[3.11]}
B.~McMahan \emph{et~al.}, ``Communication-efficient learning of deep networks
  from decentralized data,'' in \emph{Proc. Int. Conf. Artif. Intell.
  Statist.}\hskip 1em plus 0.5em minus 0.4em\relax PMLR, Apr. 2017, pp.
  1273--1282.

\bibitem{[3]}
Q.~Yang \emph{et~al.}, ``Federated machine learning: Concept and
  applications,'' \emph{ACM Trans. Intell. Syst. and Technol. (TIST)}, vol.~10,
  no.~2, Jan. 2019.

\bibitem{[9.14]}
J.~Park, S.~Samarakoon, M.~Bennis, and M.~Debbah, ``Wireless network
  intelligence at the edge,'' \emph{Proc. IEEE}, vol. 107, no.~11, pp.
  2204--2239, Oct. 2019.

\bibitem{[2.1]}
\BIBentryALTinterwordspacing
P.~Kairouz \emph{et~al.}, ``Advances and open problems in federated learning,''
  \emph{arXiv:1912.04977}, Dec. 2019. [Online]. Available:
  \url{http://arxiv.org/abs/1912.04977}
\BIBentrySTDinterwordspacing

\bibitem{[2.5]}
\BIBentryALTinterwordspacing
L.~Lyu, H.~Yu, and Q.~Yang, ``Threats to federated learning: A survey,''
  \emph{arXiv:2003.02133}, Mar. 2020. [Online]. Available:
  \url{https://arxiv.org/abs/2003.02133}
\BIBentrySTDinterwordspacing

\bibitem{[2.6]}
T.~Li \emph{et~al.}, ``Federated learning: Challenges, methods, and future
  directions,'' \emph{IEEE Signal Process.}, vol.~37, no.~3, pp. 50--60, May
  2020.

\bibitem{[2.7]}
Z.~Du \emph{et~al.}, ``Federated learning for vehicular internet of things:
  Recent advances and open issues,'' \emph{IEEE Open J. Comput. Soc.}, vol.~1,
  pp. 45--61, May 2020.

\bibitem{[2.9]}
M.~Aledhari \emph{et~al.}, ``Federated learning: {A} survey on enabling
  technologies, protocols, and applications,'' \emph{{IEEE} Access}, vol.~8,
  pp. 140\,699--140\,725, July 2020.

\bibitem{[2.4]}
V.~Kulkarni, M.~Kulkarni, and A.~Pant, ``Survey of personalization techniques
  for federated learning,'' in \emph{Proc. Fourth World Conf. Smart Trends
  Syst., Sec. and Sustain. ({WorldS}4)}, July 2020.

\bibitem{[2.3]}
W.~Yang \emph{et~al.}, ``Federated learning in mobile edge networks: {A}
  comprehensive survey,'' \emph{IEEE Commun. Surv. Tuts.}, vol.~22, no.~3, pp.
  2031--2063, July-Sept. 2020.

\bibitem{[10.2]}
M.~Chen \emph{et~al.}, ``Wireless communications for collaborative federated
  learning,'' \emph{IEEE Communi. Mag.}, vol.~58, no.~12, pp. 48--54, Dec.
  2020.

\bibitem{[2.11]}
\BIBentryALTinterwordspacing
Q.~Li \emph{et~al.}, ``A survey on federated learning systems: vision, hype and
  reality for data privacy and protection,'' \emph{arXiv:1907.09693}, Jan.
  2021. [Online]. Available: \url{http://arxiv.org/abs/1907.09693}
\BIBentrySTDinterwordspacing

\bibitem{[2.12]}
O.~A. {Wahab} \emph{et~al.}, ``Federated machine learning: Survey, multi-level
  classification, desirable criteria and future directions in communication and
  networking systems,'' \emph{IEEE Commun. Surv. Tuts.}, pp. 1--1, Feb. 2021.

\bibitem{[2.2]}
S.~Abdulrahman \emph{et~al.}, ``A survey on federated learning: The journey
  from centralized to distributed on-site learning and beyond,'' \emph{IEEE
  Internet Things J.}, vol.~8, no.~7, pp. 5476--5497, Apr. 2021.

\bibitem{[10.1]}
L.~U. Khan \emph{et~al.}, ``Federated learning for internet of things: Recent
  advances, taxonomy, and open challenges,'' \emph{IEEE Commun. Surv. \&
  Tuts.}, June 2021.

\bibitem{[10.3]}
Z.~Yang \emph{et~al.}, ``Federated learning for {6G}: Applications, challenges,
  and opportunities,'' \emph{Eng.}, Dec. 2021.

\bibitem{[11.1]}
A.~Z. Tan \emph{et~al.}, ``Towards personalized federated learning,''
  \emph{IEEE Trans. Neural Netw. Learn. Syst.}, Mar. 2022.

\bibitem{[11.2]}
B.~Ghimire and D.~B. Rawat, ``Recent advances on federated learning for
  cybersecurity and cybersecurity for federated learning for internet of
  things,'' \emph{IEEE Internet Things J.}, June 2022.

\bibitem{[3.3]}
M.~Hao \emph{et~al.}, ``Efficient and privacy-enhanced federated learning for
  industrial artificial intelligence,'' \emph{IEEE Trans. on Ind. Informat.},
  vol.~16, no.~10, pp. 6532--6542, Oct. 2019.

\bibitem{[48]}
M.~Salehi \emph{et~al.}, ``Hierarchical federated learning across heterogeneous
  cellular networks,'' in \emph{Proc. {IEEE} Int. Conf. Acoust. Speech and
  Signal Process. (ICASSP), Barcelona, Spain}, May 2020, pp. 8866--8870.

\bibitem{[9.1]}
Y.~Luo and J.~Wang, ``Technical introduction of wireless mesh network,''
  \emph{Monitoring and Control (ANMC) Cooperate: Xi'an Technological University
  (CHINA) West Virginia University (USA) Huddersfield University of UK (UK)},
  p.~73, June 2021.

\bibitem{[1.5]}
\BIBentryALTinterwordspacing
H.~B. McMahan \emph{et~al.}, ``Federated learning of deep networks using model
  averaging,'' \emph{arXiv:1602.05629}, Feb. 2016. [Online]. Available:
  \url{http://arxiv.org/abs/1602.05629}
\BIBentrySTDinterwordspacing

\bibitem{[5]}
\BIBentryALTinterwordspacing
K.~Bonawitz \emph{et~al.}, ``Towards federated learning at scale: System
  design,'' \emph{arXiv:1902.01046}, Feb. 2019. [Online]. Available:
  \url{https://arxiv.org/abs/1902.01046}
\BIBentrySTDinterwordspacing

\bibitem{[3.12]}
\BIBentryALTinterwordspacing
F.~Lai \emph{et~al.}, ``Oort: Informed participant selection for scalable
  federated learning,'' \emph{arXiv preprint arXiv:2010.06081}, Oct. 2020.
  [Online]. Available: \url{https://arxiv.org/abs/2010.06081}
\BIBentrySTDinterwordspacing

\bibitem{[4.13]}
M.~R. Sprague \emph{et~al.}, ``Asynchronous federated learning for geospatial
  applications,'' in \emph{Proc. Conf. Mach. Learn. Knowl. Discov.
  Databases}.\hskip 1em plus 0.5em minus 0.4em\relax Springer, Nov. 2018, pp.
  21--28.

\bibitem{[12.1]}
J.~Nguyen \emph{et~al.}, ``Federated learning with buffered asynchronous
  aggregation,'' in \emph{Int. Conf. Artif. Intell. Stat.}\hskip 1em plus 0.5em
  minus 0.4em\relax PMLR, May 2022, pp. 3581--3607.

\bibitem{[12.2]}
M.~Chen, B.~Mao, and T.~Ma, ``{FedSA: A staleness-aware asynchronous federated
  learning algorithm with non-IID data},'' \emph{Future Generation Computer
  Systems}, vol. 120, pp. 1--12, July 2021.

\bibitem{[12.3]}
W.~Wu \emph{et~al.}, ``{SAFA}: A semi-asynchronous protocol for fast federated
  learning with low overhead,'' \emph{IEEE Trans. Computers}, vol.~70, no.~5,
  pp. 655--668, May 2020.

\bibitem{[3.15]}
\BIBentryALTinterwordspacing
S.~Ruder, ``An overview of gradient descent optimization algorithms,''
  \emph{arXiv preprint arXiv:1609.04747}, Sept. 2016. [Online]. Available:
  \url{https://arxiv.org/abs/1609.04747}
\BIBentrySTDinterwordspacing

\bibitem{[3.18]}
\BIBentryALTinterwordspacing
T.~Li \emph{et~al.}, ``Federated optimization in heterogeneous networks,''
  \emph{arXiv preprint arXiv:1812.06127}, 2018. [Online]. Available:
  \url{https://arxiv.org/abs/1812.06127}
\BIBentrySTDinterwordspacing

\bibitem{[3.19]}
\BIBentryALTinterwordspacing
R.~Pathak and M.~J. Wainwright, ``{FedSplit}: An algorithmic framework for fast
  federated optimization,'' \emph{arXiv preprint arXiv:2005.05238}, May 2020.
  [Online]. Available: \url{https://arxiv.org/abs/2005.05238}
\BIBentrySTDinterwordspacing

\bibitem{[3.21]}
\BIBentryALTinterwordspacing
S.~Reddi \emph{et~al.}, ``Adaptive federated optimization,'' \emph{arXiv
  preprint arXiv:2003.00295}, Feb. 2020. [Online]. Available:
  \url{https://arxiv.org/abs/2003.00295}
\BIBentrySTDinterwordspacing

\bibitem{[3.20]}
\BIBentryALTinterwordspacing
D.~Basu \emph{et~al.}, ``{Qsparse-local-SGD}: Distributed {SGD} with
  quantization, sparsification, and local computations,'' \emph{arXiv preprint
  arXiv:1906.02367}, June 2019. [Online]. Available:
  \url{https://arxiv.org/abs/1906.02367}
\BIBentrySTDinterwordspacing

\bibitem{[11.3]}
J.~Hamer, M.~Mohri, and A.~T. Suresh, ``Fedboost: A communication-efficient
  algorithm for federated learning,'' in \emph{Proc. Int. Conf. Mach.
  Learn.}\hskip 1em plus 0.5em minus 0.4em\relax PMLR, Nov. 2020, pp.
  3973--3983.

\bibitem{[11.4]}
\BIBentryALTinterwordspacing
M.~Al-Quraan \emph{et~al.}, ``Fedtrees: A novel computation-communication
  efficient federated learning framework investigated in smart grids,''
  \emph{arXiv preprint arXiv:2210.00060}, Oct. 2022. [Online]. Available:
  \url{http://arxiv.org/abs/2210.00060}
\BIBentrySTDinterwordspacing

\bibitem{[3.22]}
\BIBentryALTinterwordspacing
D.~\-Lhuissier, ``Etsi - multi-access edge computing - standards for {MEC},''
  2021. [Online]. Available:
  \url{https://www.etsi.org/technologies/multi-access-edge-computing?jjj=1622661275132}
\BIBentrySTDinterwordspacing

\bibitem{[3.32]}
C.~S. Wright, ``Bitcoin: A peer-to-peer electronic cash system,''
  \emph{Available at SSRN 3440802}, Oct. 2008.

\bibitem{[5.2]}
K.~Toyoda and A.~N. Zhang, ``Mechanism design for an incentive-aware
  blockchain-enabled federated learning platform,'' in \emph{Proc. IEEE Int.
  Conf. Big Data (Big Data), Los Angeles, CA, USA}, Dec. 2019, pp. 395--403.

\bibitem{[7.1]}
D.~C. Nguyen \emph{et~al.}, ``Federated learning meets blockchain in edge
  computing: Opportunities and challenges,'' \emph{IEEE Internet Things J.},
  Apr. 2021.

\bibitem{[5.3]}
Y.~Zhao \emph{et~al.}, ``Privacy-preserving blockchain-based federated learning
  for {IoT} devices,'' \emph{IEEE Internet Things J.}, vol.~8, no.~3, pp.
  1817--1829, Aug. 2020.

\bibitem{[11.5]}
M.~Aloqaily, I.~Al~Ridhawi, and M.~Guizani, ``Energy-aware blockchain and
  federated learning-supported vehicular networks,'' \emph{IEEE Trans. Intell.
  Transp. Syst.}, Aug. 2021.

\bibitem{[47]}
Y.~Lu \emph{et~al.}, ``Blockchain and federated learning for {5G} beyond,''
  \emph{IEEE Netw.}, vol.~35, no.~1, pp. 219--225, Jan./Feb. 2021.

\bibitem{[3.36]}
X.~Foukas, G.~Patounas, A.~Elmokashfi, and M.~K. Marina, ``Network slicing in
  {5G}: Survey and challenges,'' \emph{IEEE Commun. Mag.}, vol.~55, no.~5, pp.
  94--100, May 2017.

\bibitem{[63]}
B.~Brik and A.~Ksentini, ``On predicting service-oriented network slices
  performances in {5G}: {A} federated learning approach,'' in \emph{Proc. 45th
  {IEEE} Conf. Local Comput. Netw. (LCN), Sydney, NSW, Australia}, Nov. 2020,
  pp. 164--171.

\bibitem{[11.7]}
H.~P. Phyu, D.~Naboulsi, and R.~Stanica, ``Mobile traffic forecasting for
  network slices: A federated-learning approach,'' in \emph{Proc. 33rd Int.
  Symp. Pers. Indoor Mobile Radio Commun. (PIMRC)}.\hskip 1em plus 0.5em minus
  0.4em\relax IEEE, Sept. 2022.

\bibitem{[11.8]}
S.~Messaoud \emph{et~al.}, ``{Deep federated Q-learning-based network slicing
  for industrial IoT},'' \emph{IEEE Trans. Ind. Inform.}, vol.~17, no.~8, pp.
  5572--5582, Oct. 2020.

\bibitem{[65]}
Y.~Liu \emph{et~al.}, ``Device association for {RAN} slicing based on hybrid
  federated deep reinforcement learning,'' \emph{IEEE Trans. Veh. Technol},
  vol.~69, no.~12, pp. 15\,731--15\,745, Dec. 2020.

\bibitem{[4.16]}
A.~Aissioui \emph{et~al.}, ``On enabling {5G} automotive systems using follow
  me edge-cloud concept,'' \emph{IEEE Trans. Veh. Technol.}, vol.~67, no.~6,
  pp. 5302--5316, Feb. 2018.

\bibitem{[50]}
T.~Subramanya and R.~Riggio, ``Centralized and federated learning for
  predictive {VNF} autoscaling in multi-domain {5G} networks and beyond,''
  \emph{IEEE Trans. Netw. Service Manag.}, vol.~18, no.~1, pp. 63--78, Mar.
  2021.

\bibitem{[56]}
Y.~Wei \emph{et~al.}, ``Federated learning empowered end-edge-cloud cooperation
  for {5G HetNet} security,'' \emph{IEEE Netw.}, vol.~35, no.~2, pp. 88--94,
  Mar./Apr. 2021.

\bibitem{[49]}
\BIBentryALTinterwordspacing
S.~Jere \emph{et~al.}, ``Federated learning in mobile edge computing: An
  edge-learning perspective for beyond {5G},'' \emph{arXiv:2007.08030}, July
  2020. [Online]. Available: \url{https://arxiv.org/abs/2007.08030}
\BIBentrySTDinterwordspacing

\bibitem{[11.9]}
A.~M. Elbir and S.~Coleri, ``Federated learning for channel estimation in
  conventional and {RIS}-assisted massive {MIMO},'' \emph{IEEE Trans. Wireless
  Commun.}, Nov. 2021.

\bibitem{[11.10]}
T.~Vu \emph{et~al.}, ``Energy-efficient massive {MIMO} for serving multiple
  federated learning groups,'' in \emph{Proc. Global Commun. Conf. (GLOBECOM),
  Madrid, Spain}.\hskip 1em plus 0.5em minus 0.4em\relax IEEE, Dec. 2021, pp.
  1--6.

\bibitem{[11.11]}
D.~Yu \emph{et~al.}, ``Optimizing over-the-air computation in {IRS-aided C-RAN}
  systems,'' in \emph{21st Int. Workshop Signal Process. Adv. Wireless Commun.
  (SPAWC), Atlanta, GA, USA}.\hskip 1em plus 0.5em minus 0.4em\relax IEEE, May
  2020, pp. 1--5.

\bibitem{[11.12]}
S.~Manzoor \emph{et~al.}, ``Federated learning empowered mobility-aware
  proactive content offloading framework for fog radio access networks,''
  \emph{Future Gener. Comput. Syst.}, vol. 133, pp. 307--319, Aug. 2022.

\bibitem{[11.13]}
\BIBentryALTinterwordspacing
Z.~Ji and Z.~Qin, ``Federated learning for distributed energy-efficient
  resource allocation,'' \emph{arXiv preprint arXiv:2204.09602}, Apr. 2022.
  [Online]. Available: \url{http://arxiv.org/abs/2204.09602}
\BIBentrySTDinterwordspacing

\bibitem{[4.22]}
A.~Hammoud \emph{et~al.}, ``{AI}, blockchain, and vehicular edge computing for
  smart and secure {IoV}: Challenges and directions,'' \emph{IEEE Internet
  Things Mag.}, vol.~3, no.~2, pp. 68--73, June 2020.

\bibitem{[34]}
S.~R. Pokhrel and J.~Choi, ``Federated learning with blockchain for autonomous
  vehicles: Analysis and design challenges,'' \emph{IEEE Trans. Commun.},
  vol.~68, no.~8, pp. 4734--4746, Aug. 2020.

\bibitem{[9.3]}
S.~{S}amarakoon, M.~Bennis, W.~Saad, and M.~Debbah, ``Distributed federated
  learning for ultra-reliable low-latency vehicular communications,''
  \emph{IEEE Trans. Commun.}, vol.~68, no.~2, pp. 1146--1159, Nov. 2019.

\bibitem{[33]}
D.~Ye \emph{et~al.}, ``Federated learning in vehicular edge computing: {A}
  selective model aggregation approach,'' \emph{{IEEE} Access}, vol.~8, pp.
  23\,920--23\,935, Jan. 2020.

\bibitem{[36]}
Y.~M. Saputra \emph{et~al.}, ``Federated learning meets contract theory:
  Economic-efficiency framework for electric vehicle networks,'' \emph{IEEE
  Trans. Mobile Comput.}, pp. 1--1, Dec. 2020.

\bibitem{[44]}
B.~Qolomany \emph{et~al.}, ``Particle swarm optimized federated learning for
  industrial {IoT} and smart city services,'' in \emph{Proc. IEEE Global
  Commun. Conf. (GLOBECOM)}, Dec. 2020, pp. 1--6.

\bibitem{[4.26]}
\BIBentryALTinterwordspacing
A.~Rovira-Sugranes \emph{et~al.}, ``A review of {AI}-enabled routing protocols
  for {UAV} networks: Trends, challenges, and future outlook,'' \emph{arXiv
  preprint arXiv:2104.01283}, Apr. 2021. [Online]. Available:
  \url{http://arxiv.org/abs/2104.01283}
\BIBentrySTDinterwordspacing

\bibitem{[37]}
Y.~Wang \emph{et~al.}, ``Learning in the air: Secure federated learning for
  {UAV}-assisted crowdsensing,'' \emph{IEEE Trans. Netw. Sci. Eng.}, pp. 1--1,
  Aug. 2021.

\bibitem{[38]}
Y.~Liu \emph{et~al.}, ``Federated learning in the sky: Aerial-ground air
  quality sensing framework with {UAV} swarms,'' \emph{IEEE Internet Things
  J.}, Sept. 2020.

\bibitem{[42]}
\BIBentryALTinterwordspacing
S.~Tang \emph{et~al.}, ``Battery-constrained federated edge learning in
  {UAV-enabled} {IoT} for {B5G/6G} networks,'' \emph{arXiv:2101.12472}, Jan.
  2021. [Online]. Available: \url{https://arxiv.org/abs/2101.12472}
\BIBentrySTDinterwordspacing

\bibitem{[39]}
H.~Zhang and L.~Hanzo, ``Federated learning assisted multi-{UAV} networks,''
  \emph{IEEE Trans. Veh. Technol.}, vol.~69, no.~11, pp. 14\,104--14\,109, Nov.
  2020.

\bibitem{[4.29]}
W.~Ni \emph{et~al.}, ``Optimal transmission control and learning-based
  trajectory design for {UAV}-assisted detection and communication,'' in
  \emph{Proc. IEEE 31st Ann. Int. Symp. Pers., Indoor Mobile Radio Commun.
  (PIMRC), London, UK}, Oct. 2020, pp. 1--6.

\bibitem{[4.30]}
H.~A.~B. Salameh \emph{et~al.}, ``Jamming-aware simultaneous multi-channel
  decisions for opportunistic access in delay-critical {IoT}-based sensor
  networks,'' \emph{IEEE Sensors J.}, vol.~22, no.~3, pp. 2889--2898, Dec.
  2021.

\bibitem{[40]}
N.~I. Mowla \emph{et~al.}, ``Federated learning-based cognitive detection of
  jamming attack in flying {Ad-Hoc} network,'' \emph{{IEEE} Access}, vol.~8,
  pp. 4338--4350, Dec. 2019.

\bibitem{[4.31]}
E.~Basar \emph{et~al.}, ``Wireless communications through reconfigurable
  intelligent surfaces,'' \emph{IEEE Access}, vol.~7, pp. 116\,753--116\,773,
  Aug. 2019.

\bibitem{[55]}
K.~Yang \emph{et~al.}, ``Federated machine learning for intelligent {IoT} via
  reconfigurable intelligent surface,'' \emph{IEEE Netw.}, vol.~34, no.~5, pp.
  16--22, Sept. 2020.

\bibitem{[80]}
H.~Liu, X.~Yuan, and Y.-J.~A. Zhang, ``Reconfigurable intelligent surface
  enabled federated learning: A unified communication-learning design
  approach,'' \emph{IEEE Trans. Wireless Commun.}, June 2021.

\bibitem{[82]}
\BIBentryALTinterwordspacing
H.~Liu, X.~Yuan, and Y.~A. Zhang, ``{CSIT-Free} federated edge learning via
  reconfigurable intelligent surface,'' \emph{arXiv:2102.10749}, Feb. 2021.
  [Online]. Available: \url{https://arxiv.org/abs/2102.10749}
\BIBentrySTDinterwordspacing

\bibitem{[81]}
L.~Li \emph{et~al.}, ``Enhanced reconfigurable intelligent surface assisted
  mm{W}ave communication: A federated learning approach,'' \emph{Chin.
  Commun.}, vol.~17, no.~10, pp. 115--128, Oct. 2020.

\bibitem{[4.33]}
Y.~Shaikh \emph{et~al.}, ``Survey of smart healthcare systems using internet of
  things {IoT},'' in \emph{Proc. IEEE Int. Conf. Commun. Comput. Internet
  Things (IC3IoT), Chennai, India}, Feb. 2018, pp. 508--513.

\bibitem{[4.34]}
L.~Romeo \emph{et~al.}, ``Internet of robotic things in smart domains:
  Applications and challenges,'' \emph{Sensors}, vol.~20, no.~12, p. 3355, Jan.
  2020.

\bibitem{[52]}
I.~Mohammed \emph{et~al.}, ``Budgeted online selection of candidate {IoT}
  clients to participate in federated learning,'' \emph{IEEE Internet Things
  J.}, vol.~8, no.~7, pp. 5938 -- 5952, Apr. 2020.

\bibitem{[54]}
Y.~Liu \emph{et~al.}, ``Deep anomaly detection for time-series data in
  industrial {IoT}: {A} communication-efficient on-device federated learning
  approach,'' \emph{IEEE Internet Things J.}, vol.~8, no.~8, pp. 6348 -- 6358,
  Apr. 2020.

\bibitem{[4.40]}
Y.~Chen \emph{et~al.}, ``Fedhealth: A federated transfer learning framework for
  wearable healthcare,'' \emph{IEEE Intell. Syst.}, vol.~35, no.~4, pp. 83--93,
  Apr. 2020.

\bibitem{[45]}
\BIBentryALTinterwordspacing
B.~Liu \emph{et~al.}, ``Experiments of federated learning for {COVID-19} chest
  {X-ray} images,'' \emph{arXiv:2007.05592}, July 2020. [Online]. Available:
  \url{https://arxiv.org/abs/2007.05592}
\BIBentrySTDinterwordspacing

\bibitem{[46]}
W.~Yang \emph{et~al.}, ``{FFD:} {A} federated learning based method for credit
  card fraud detection,'' in \emph{Proc. Big Data 8th Int. Congr., Services
  Conf. Federation, (SCF), San Diego, CA, USA}, K.~Chen, S.~Seshadri, and
  L.~Zhang, Eds., vol. 11514, June 2019, pp. 18--32.

\bibitem{[3.50]}
J.~Singh \emph{et~al.}, ``Micro-{LED} as a promising candidate for high-speed
  visible light communication,'' \emph{Appl. Sci.}, vol.~10, no.~20, p. 7384,
  Jan. 2020.

\bibitem{[3.51]}
S.~Idris \emph{et~al.}, ``Visible light communication: A potential {5G} and
  beyond communication technology,'' in \emph{Proc. 15th Int. Conf. Electron.,
  Comput. Comput. (ICECCO), Abuja, Nigeria}, Dec. 2019, pp. 1--6.

\bibitem{[4.58]}
J.~Zhang \emph{et~al.}, ``Cell-free massive {MIMO}: A new next-generation
  paradigm,'' \emph{IEEE Access}, vol.~7, pp. 99\,878--99\,888, July 2019.

\bibitem{[3.56]}
T.~L. Marzetta, \emph{Fundamentals of massive {MIMO}}.\hskip 1em plus 0.5em
  minus 0.4em\relax Cambridge University Press, Nov. 2016.

\bibitem{[7.7]}
T.~Tozer and D.~Grace, ``High-altitude platforms for wireless communications,''
  \emph{Electron. \& Commun. Eng. J.}, vol.~13, no.~3, pp. 127--137, June 2001.

\bibitem{[7.2]}
L.~Lei \emph{et~al.}, ``Beam illumination pattern design in satellite networks:
  Learning and optimization for efficient beam hopping,'' \emph{IEEE Access},
  vol.~8, pp. 136\,655--136\,667, July 2020.

\bibitem{[3.57]}
M.~Azmat and S.~Kummer, ``Potential applications of unmanned ground and aerial
  vehicles to mitigate challenges of transport and logistics-related critical
  success factors in the humanitarian supply chain,'' \emph{Asian J. Sustain.
  Social Responsib.}, vol.~5, no.~1, pp. 1--22, Dec. 2020.

\bibitem{[11.14]}
Q.~Lan \emph{et~al.}, ``What is semantic communication? a view on conveying
  meaning in the era of machine intelligence,'' \emph{J. Commun. Inform.
  Netw.}, vol.~6, no.~4, pp. 336--371, Dec. 2021.

\bibitem{[11.15]}
V.~Balasubramanian \emph{et~al.}, ``{Intelligent resource management at the
  edge for ubiquitous IoT: an SDN-based federated learning approach},''
  \emph{IEEE netw.}, vol.~35, no.~5, pp. 114--121, Nov. 2021.

\bibitem{[72]}
\BIBentryALTinterwordspacing
A.~G. Roy \emph{et~al.}, ``{BrainTorrent}: {A} peer-to-peer environment for
  decentralized federated learning,'' \emph{arXiv:1905.06731}, May 2019.
  [Online]. Available: \url{http://arxiv.org/abs/1905.06731}
\BIBentrySTDinterwordspacing

\bibitem{[74]}
T.~Nishio and R.~Yonetani, ``Client selection for federated learning with
  heterogeneous resources in mobile edge,'' \emph{IEEE Int. Conf. Commun.
  (ICC)}, pp. 1--7, May 2019.

\bibitem{[4.45]}
Y.~Zhan \emph{et~al.}, ``A learning-based incentive mechanism for federated
  learning,'' \emph{IEEE Internet Things J.}, vol.~7, no.~7, pp. 6360--6368,
  Jan. 2020.

\bibitem{[4.46]}
J.~Kang and Xothers, ``Incentive mechanism for reliable federated learning: A
  joint optimization approach to combining reputation and contract theory,''
  \emph{IEEE Internet Things J.}, vol.~6, no.~6, pp. 10\,700--10\,714, Sept.
  2019.

\bibitem{[4.47]}
\BIBentryALTinterwordspacing
H.~Yang, M.~Fang, and J.~Liu, ``Achieving linear speedup with partial worker
  participation in non-{IID} federated learning,'' \emph{arXiv preprint
  arXiv:2101.11203}, Jan. 2021. [Online]. Available:
  \url{http://arxiv.org/abs/2101.11203}
\BIBentrySTDinterwordspacing

\bibitem{[51]}
Q.~Wu, K.~He, and X.~Chen, ``Personalized federated learning for intelligent
  {IoT} applications: A cloud-edge based framework,'' \emph{IEEE Open J.
  Comput. Soc.}, vol.~1, pp. 35--44, Feb. 2020.

\bibitem{[12.4]}
\BIBentryALTinterwordspacing
Y.~Ezzeldin \emph{et~al.}, ``{FairFed}: Enabling group fairness in federated
  learning,'' \emph{arXiv preprint arXiv:2110.00857}, Oct. 2021. [Online].
  Available: \url{http://arxiv.org/abs/2110.00857}
\BIBentrySTDinterwordspacing

\bibitem{[12.5]}
T.~Huang \emph{et~al.}, ``An efficiency-boosting client selection scheme for
  federated learning with fairness guarantee,'' \emph{IEEE Trans. Parall.
  Distrib. Syst.}, vol.~32, no.~7, pp. 1552--1564, Nov. 2020.

\bibitem{[5.4]}
\BIBentryALTinterwordspacing
S.~Augenstein \emph{et~al.}, ``Generative models for effective {ML} on private,
  decentralized datasets,'' \emph{arXiv preprint arXiv:1911.06679}, Nov. 2019.
  [Online]. Available: \url{http://arxiv.org/abs/1911.06679}
\BIBentrySTDinterwordspacing

\bibitem{[43]}
A.~Albaseer \emph{et~al.}, ``Exploiting unlabeled data in smart cities using
  federated edge learning,'' in \emph{Proc. 16th Int. Wireless Commun. Mobile
  Comput. Conf. (IWCMC), Limassol, Cyprus}, June 2020, pp. 1666--1671.

\bibitem{[3.28]}
K.~Yang \emph{et~al.}, ``Federated learning via over-the-air computation,''
  \emph{IEEE Trans. Wireless Commun.}, vol.~19, no.~3, pp. 2022--2035, Jan.
  2020.

\bibitem{[3.29]}
G.~Zhu, Y.~Wang, and K.~Huang, ``Broadband analog aggregation for low-latency
  federated edge learning,'' \emph{IEEE Trans. Wireless Commun.}, vol.~19,
  no.~1, pp. 491--506, Oct. 2019.

\bibitem{[3.30]}
G.~Zhu \emph{et~al.}, ``One-bit over-the-air aggregation for
  communication-efficient federated edge learning: Design and convergence
  analysis,'' \emph{IEEE Trans. Wireless Commun.}, vol.~20, no.~3, pp.
  2120--2135, Nov. 2020.

\bibitem{[4.55]}
X.~Yao, C.~Huang, and L.~Sun, ``Two-stream federated learning: Reduce the
  communication costs,'' in \emph{Proc. IEEE Vis. Commun. Imag. Process.
  (VCIP), Taichung, Taiwan}, Dec. 2018, pp. 1--4.

\bibitem{[73]}
\BIBentryALTinterwordspacing
Y.~Lin \emph{et~al.}, ``Deep gradient compression: Reducing the communication
  bandwidth for distributed training,'' \emph{arXiv:1712.01887}, Dec. 2017.
  [Online]. Available: \url{https://arxiv.org/abs/1712.01887}
\BIBentrySTDinterwordspacing

\bibitem{[4.52]}
H.~Sun, X.~Ma, and R.~Q. Hu, ``Adaptive federated learning with gradient
  compression in uplink {NOMA},'' \emph{IEEE Trans. Veh. Technol.}, vol.~69,
  no.~12, pp. 16\,325--16\,329, Sept. 2020.

\bibitem{[3.39]}
M.~Chen \emph{et~al.}, ``Convergence time optimization for federated learning
  over wireless networks,'' \emph{IEEE Trans. Wireless Commun.}, vol.~20,
  no.~4, pp. 2457--2471, Dec. 2020.

\bibitem{[3.40]}
\BIBentryALTinterwordspacing
T.~Huang \emph{et~al.}, ``Stochastic client selection for federated learning
  with volatile clients,'' \emph{arXiv preprint arXiv:2011.08756}, Nov. 2020.
  [Online]. Available: \url{https://arxiv.org/abs/2011.08756}
\BIBentrySTDinterwordspacing

\bibitem{[3.38]}
W.~Xia, W.~Wen, K.-K. Wong, T.~Q. Quek, J.~Zhang, and H.~Zhu,
  ``Federated-learning-based client scheduling for low-latency wireless
  communications,'' \emph{IEEE Wireless Commun.}, vol.~28, no.~2, pp. 32--38,
  May 2021.

\bibitem{[3.47]}
L.~N. Smith, ``Cyclical learning rates for training neural networks,'' in
  \emph{Proc. IEEE winter conf. appl. comput. vision (WACV), Santa Rosa, CA,
  USA}, Mar. 2017, pp. 464--472.

\bibitem{[9.16]}
C.~Zhao \emph{et~al.}, ``Secure multi-party computation: theory, practice and
  applications,'' \emph{Inform. Sci.}, vol. 476, pp. 357--372, Feb. 2019.

\bibitem{[9.18]}
\BIBentryALTinterwordspacing
H.~B. McMahan \emph{et~al.}, ``Learning differentially private recurrent
  language models,'' \emph{arXiv preprint arXiv:1710.06963}, Oct. 2017.
  [Online]. Available: \url{http://arxiv.org/abs/1710.06963}
\BIBentrySTDinterwordspacing

\bibitem{[35]}
Y.~Lu \emph{et~al.}, ``Differentially private asynchronous federated learning
  for mobile edge computing in urban informatics,'' \emph{IEEE Trans. Ind.
  Informat.}, vol.~16, no.~3, pp. 2134--2143, Mar. 2020.

\bibitem{[76]}
C.~Zhang \emph{et~al.}, ``{BatchCrypt}: Efficient homomorphic encryption for
  cross-silo federated learning,'' in \emph{Proc. {USENIX} Annu. Techn. Conf.
  (ATC)}, July 2020, pp. 493--506.

\bibitem{[4.62]}
Y.~Sun \emph{et~al.}, ``Update or wait: How to keep your data fresh,''
  \emph{IEEE Trans. Inform. Theory}, vol.~63, no.~11, pp. 7492--7508, Aug.
  2017.

\bibitem{[4.64]}
M.~Costa, M.~Codreanu, and A.~Ephremides, ``On the age of information in status
  update systems with packet management,'' \emph{IEEE Trans. Inform. Theory},
  vol.~62, no.~4, pp. 1897--1910, Feb. 2016.

\bibitem{[4.67]}
\BIBentryALTinterwordspacing
S.~Bayhan, G.~G{\"u}r, and A.~Zubow, ``The future is unlicensed: Coexistence in
  the unlicensed spectrum for {5G},'' \emph{arXiv preprint arXiv:1801.04964},
  Jan. 2018. [Online]. Available: \url{http://arxiv.org/abs/1801.04964}
\BIBentrySTDinterwordspacing

\bibitem{[4.72]}
\BIBentryALTinterwordspacing
T.~Hospedales \emph{et~al.}, ``Meta-learning in neural networks: A survey,''
  \emph{arXiv preprint arXiv:2004.05439}, Apr. 2020. [Online]. Available:
  \url{http://arxiv.org/abs/2004.05439}
\BIBentrySTDinterwordspacing

\bibitem{[5.6]}
\BIBentryALTinterwordspacing
A.~Jaegle \emph{et~al.}, ``Perceiver: {G}eneral perception with iterative
  attention,'' \emph{arXiv preprint arXiv:2103.03206}, Mar. 2021. [Online].
  Available: \url{http://arxiv.org/abs/2103.03206}
\BIBentrySTDinterwordspacing

\bibitem{[5.7]}
A.~Vaswani \emph{et~al.}, ``Attention is all you need,'' in \emph{Proc. Adv.
  neural inform. process. syst.}, 2017, pp. 5998--6008.

\bibitem{[9.20]}
\BIBentryALTinterwordspacing
C.-J. Wu \emph{et~al.}, ``Sustainable {AI}: Environmental implications,
  challenges and opportunities,'' \emph{arXiv preprint arXiv:2111.00364}, Oct.
  2021. [Online]. Available: \url{http://arxiv.org/abs/2111.00364}
\BIBentrySTDinterwordspacing

\bibitem{[9.29]}
\BIBentryALTinterwordspacing
B.~Guler and A.~Yener, ``Sustainable federated learning,'' \emph{arXiv preprint
  arXiv:2102.11274}, Feb. 2021. [Online]. Available:
  \url{http://arxiv.org/abs/2102.11274}
\BIBentrySTDinterwordspacing

\bibitem{[9.7]}
Y.~He, X.~Zhang, and J.~Sun, ``Channel pruning for accelerating very deep
  neural networks,'' in \emph{Proc. IEEE int. conf. comput. vis. (ICCV)}, 2017,
  pp. 1389--1397.

\bibitem{[9.10]}
S.~Chen, W.~Wang, and S.~J. Pan, ``Metaquant: Learning to quantize by learning
  to penetrate non-differentiable quantization,'' \emph{Adv. Neural Inform.
  Process. Syst.}, vol.~32, pp. 3916--3926, 2019.

\bibitem{[9.13]}
H.~Qin \emph{et~al.}, ``Binary neural networks: A survey,'' \emph{Pattern
  Recognit.}, vol. 105, p. 107281, Sept. 2020.

\bibitem{[7.8]}
\BIBentryALTinterwordspacing
L.~U. Khan \emph{et~al.}, ``{Digital-Twin-Enabled 6G}: Vision, architectural
  trends, and future directions,'' \emph{arXiv preprint arXiv:2102.12169}, Feb.
  2021. [Online]. Available: \url{http://arxiv.org/abs/2102.12169}
\BIBentrySTDinterwordspacing

\bibitem{[10.4]}
H.~Sami \emph{et~al.}, ``{AI-based resource provisioning of IoE services in 6G:
  A deep reinforcement learning approach},'' \emph{IEEE Trans. Net. Service
  Manage.}, vol.~18, no.~3, pp. 3527--3540, Mar. 2021.

\bibitem{[4.81]}
L.~L. Pipino, Y.~W. Lee, and R.~Y. Wang, ``Data quality assessment,''
  \emph{Commun. ACM}, vol.~45, no.~4, pp. 211--218, Apr. 2002.

\end{thebibliography}

%\vspace{-20 mm}

\begin{IEEEbiography}[{\includegraphics[width=1in,height=1.25in,clip,keepaspectratio]{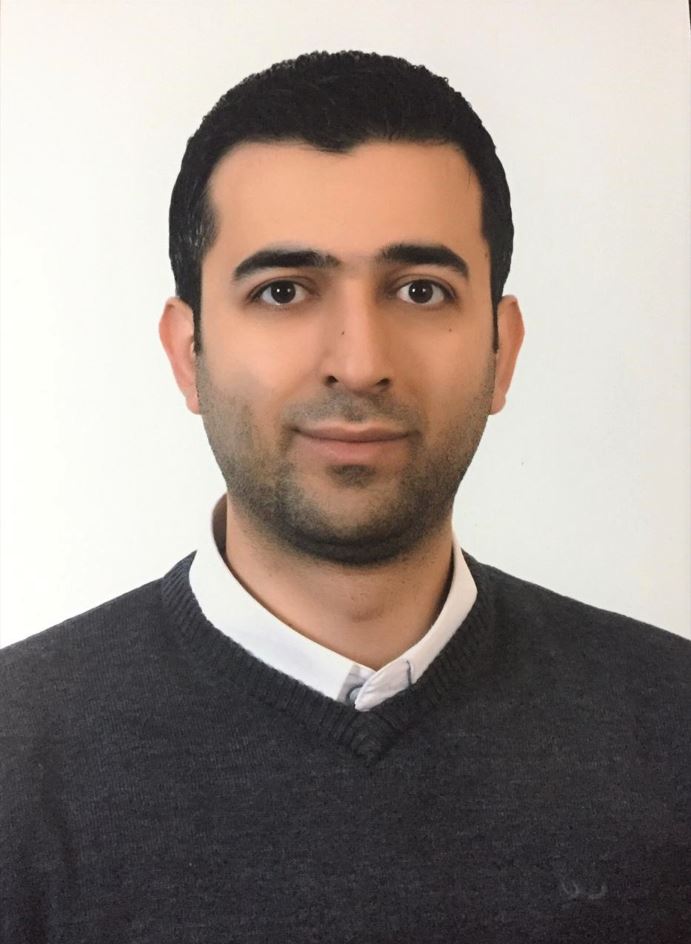}}]{Mohammad M. Al-Quraan}
received the BSc. (Honours) degree in telecommunications engineering and MSc. (Excellence) degree in wireless telecommunications engineering from Yarmouk University, Irbid, Jordan in 2011 and 2019, respectively. From 2012 to 2018, he was a senior network and telecommunications engineer at Jordan University of Science and Technology (JUST). Then until 2020, he has been the head of the network and telecommunications Department at JUST. He is currently pursuing the Ph.D. degree in electronics and electrical engineering at the University of Glasgow, Glasgow, UK. His research interests include machine learning, computer vision, cognitive radio, and beyond 5G wireless technologies.
\end{IEEEbiography}

\vskip -2\baselineskip plus -1fil

\begin{IEEEbiography}[{\includegraphics[width=1in,height=1.25in,clip,keepaspectratio]{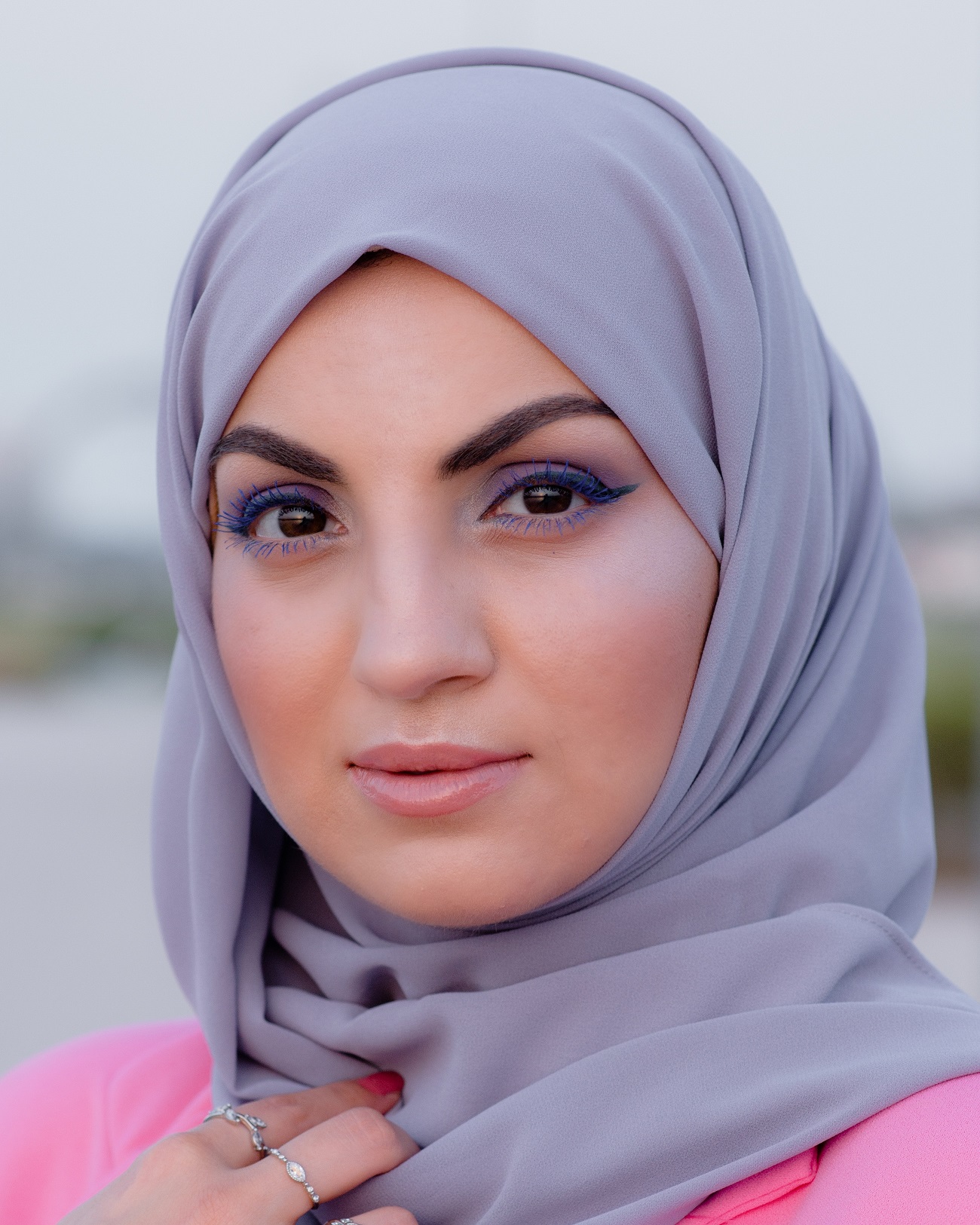}}]{Lina Mohjazi}
is an Assistant Professor (Lecturer) at the James Watt School of Engineering, University of Glasgow, U.K. She received the B.Sc. (Honours) degree from the United Arab Emirates (UAE) University, UAE, in 2008, the M.Sc. degree from Khalifa University (KU), UAE, in 2012 (Distinction), and the Ph.D. degree from the Institute for Communication Systems, University of Surrey, U.K., in 2018, all in electrical and electronic engineering. Her research interests include beyond 5G wireless technologies, physical-layer optimization and performance analysis, wireless power transfer, machine learning for future wireless systems, and reconfigurable intelligent surfaces. She is an Associate Editor for the IEEE Communications Letters and the IEEE Open Journal of the Communications Society. She is an Affiliate Member of the Mohammed bin Rashid Academy of Scientists, UAE, a Fellow of the Women's Engineering Society, and a Senior Member of the IEEE.
\end{IEEEbiography}

\vskip -2\baselineskip plus -1fil

\begin{IEEEbiography}[{\includegraphics[width=1in,height=1.25in,clip,keepaspectratio]{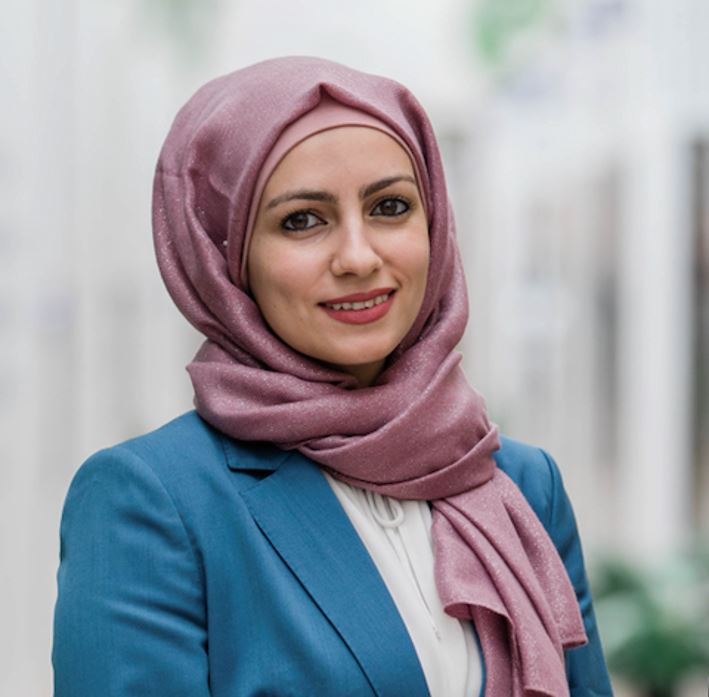}}]{Lina Bariah} received the M.Sc. and Ph.D. degrees in communications engineering from Khalifa University, Abu Dhabi, UAE, in 2015 and 2018, respectively. She was a Visiting Researcher with the Department of Systems and Computer Engineering, Carleton University, Ottawa, ON, Canada, in 2019, and an affiliate research fellow, James Watt School of Engineering, University of Glasgow, UK. She is currently a Senior Researcher at the technology Innovation institute, a visiting research scientist at Khalifa University, and an affiliate researcher in the University at Albany, USA. Dr. Bariah is a senior member of the IEEE, IEEE Communications Society, IEEE Vehicular Technology Society, and IEEE Women in Engineering. She is currently an Associate Editor for the IEEE Communication Letters, an Associate Editor for the IEEE Open Journal of the Communications Society, and an Area Editor for Physical Communication (Elsevier). She is a Guest Editor in IEEE Network Magazine, and the RS Open Journal on Innovative Communication Technologies (RS-OJICT). Dr. Bariah was a member of the technical program committee of a number of IEEE conferences, such as ICC and Globecom. She is currently organizing/chairing a number of workshops. She serves as a session chair and an active reviewer for numerous IEEE conferences and journals.
\end{IEEEbiography}

\vskip -2\baselineskip plus -1fil

\begin{IEEEbiography}[{\includegraphics[width=1in,height=1.25in,clip,keepaspectratio]{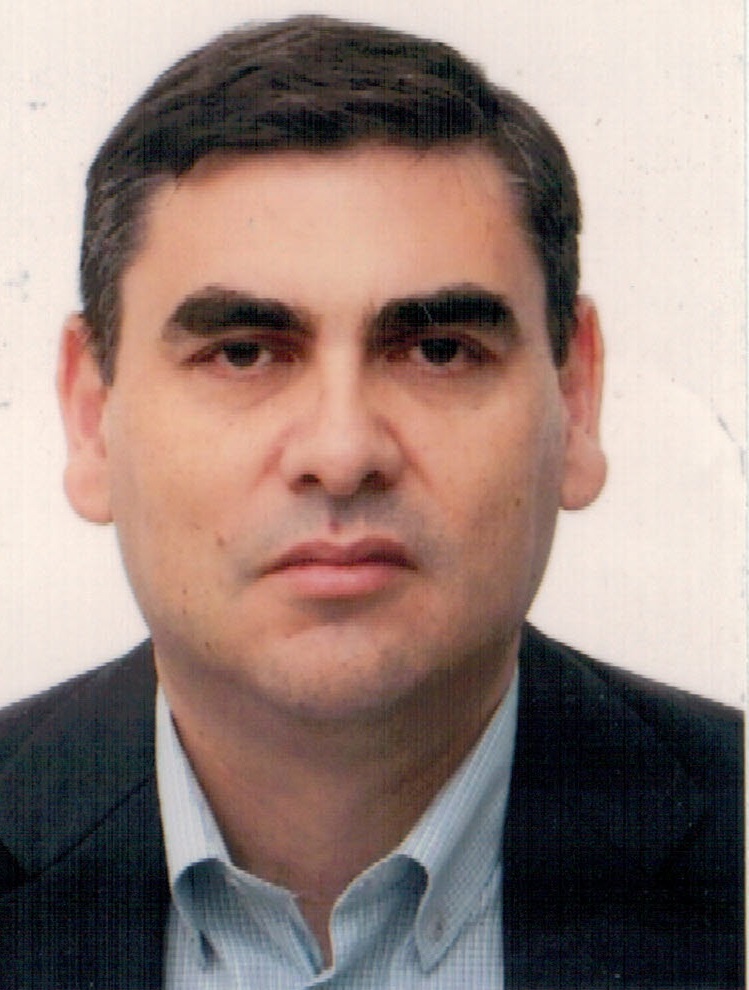}}]{Anthony Centeno}
is currently a lecturer at the University of Glasgow. He has over 25 years’ experience in industry, the UK civil service and academia. His most recent positions prior to joining Glasgow University in 2020 were as an Associate Professor at Xi’an Jiaotong Liverpool University (2017-2020) and the Malaysia-Japan International Institute of Technology (MJIIT) (2012-2017). His research is focussed on applied electromagnetism in high frequency communications, electromagnetic materials, plasmonics and electromagnetic compatibility.  He has BEng and PhD degrees from Cardiff University, is a Chartered Engineer, a member of the Institute of Physics and a member of the Institute of Engineering and Technology.  He is also a visiting honorary fellow in the Materials Department, Imperial College London.
\end{IEEEbiography}

\vskip -2\baselineskip plus -1fil

\begin{IEEEbiography}[{\includegraphics[width=1in,height=1.25in,clip,keepaspectratio]{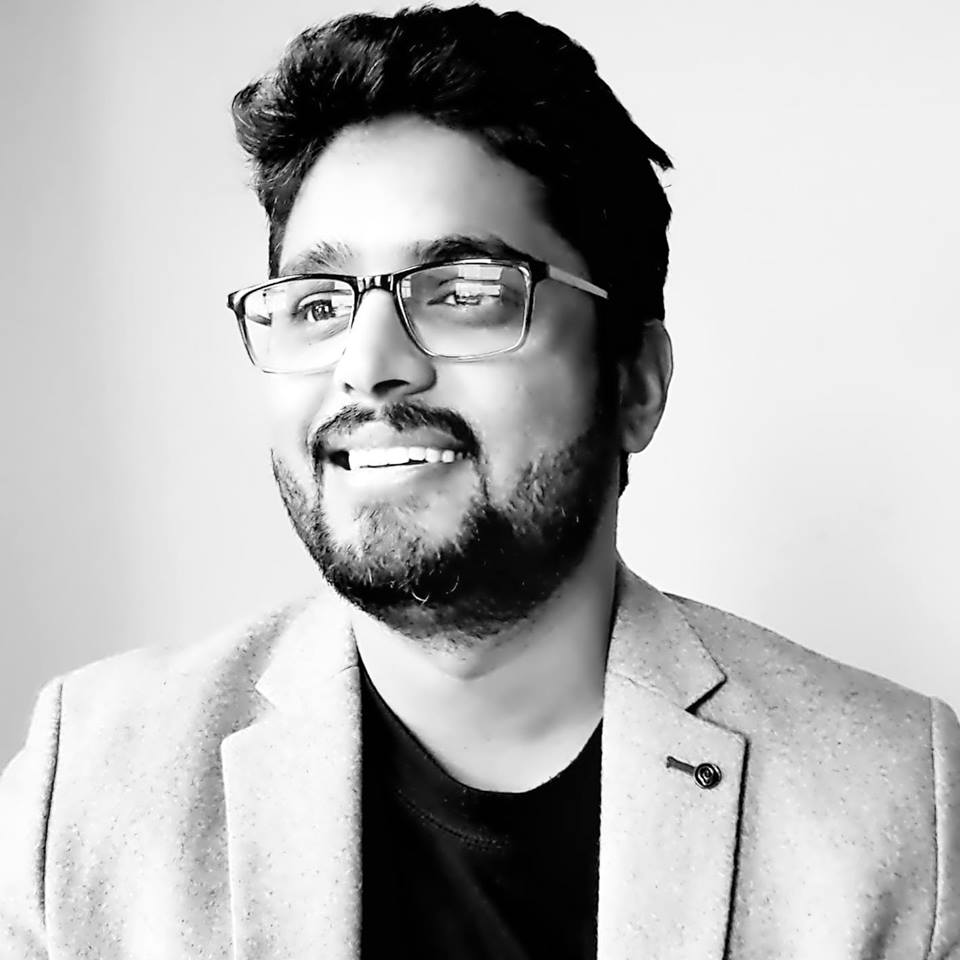}}]{Ahmed Zoha}
earned his PhD in Electrical and Electronics Engineering in July 2014 from the 6G/5GIC Centre, University of Surrey (UniS), UK, and his MSc degree in Communication Engineering from Chalmers University, Sweden. Dr Zoha has research expertise in the areas of artificial intelligence (AI) and machine learning, advanced signal processing, and state-of-the-art self-learning strategies, and he has more than 13 years of experience in designing intelligent applications and algorithms in the domain of 5G and beyond wireless communication systems, connected healthcare, internet of everything and smart energy monitoring systems. His research has been cited by national and international bodies, regulators, and the media, and he has also received two IEEE best paper awards. Dr Zoha is also a recipient of the best presentation award at GPECOMM2022 and two best paper awards at IEEE IS 2012 and IEEE SSNIP 2013 as part of his contribution to the EPSRC-funded project Reshaping Energy Demands using ICT. Dr Zoha is endorsed as a UK exceptional talent by the Royal Academy of Engineering awarded to early-career world-leading innovators and scientists. Dr Zoha has been actively involved in the organization of several IEEE and EAI conferences and served as a chair, and track chair for special sessions as well as a technical programme committee member.
\end{IEEEbiography}

\vskip -2\baselineskip plus -1fil

\begin{IEEEbiography}[{\includegraphics[width=1in,height=1.25in,clip,keepaspectratio]{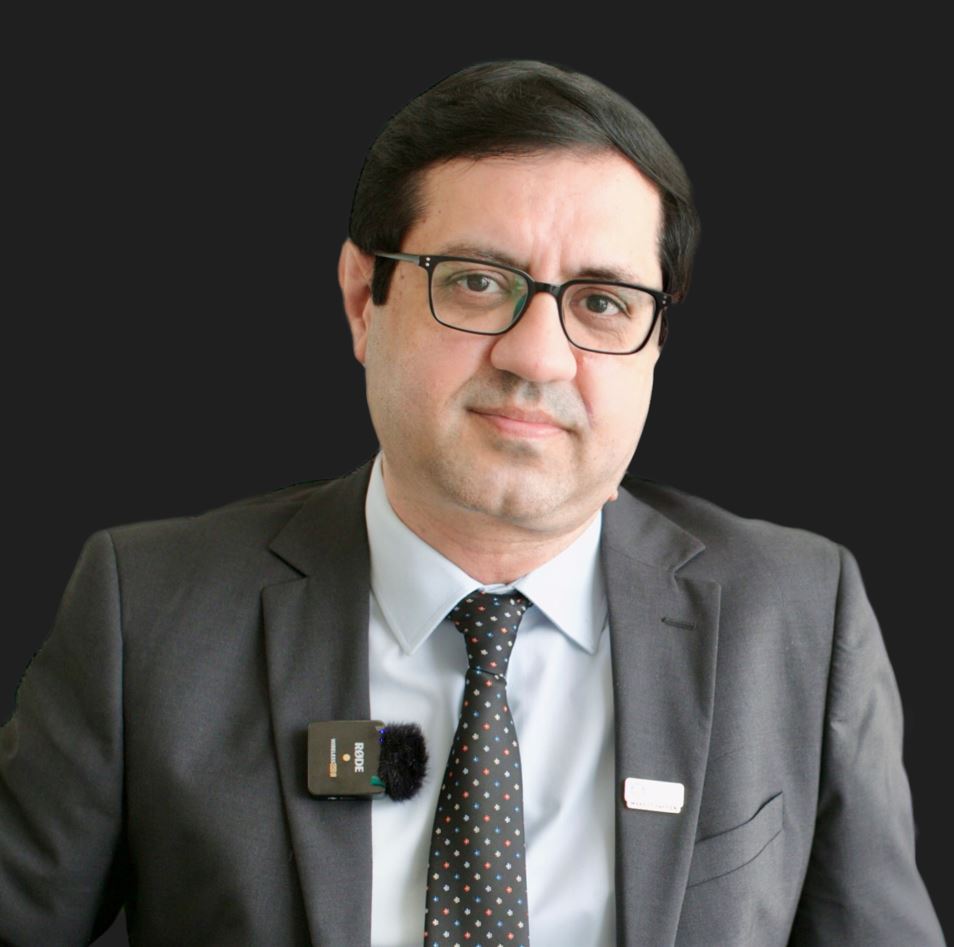}}]{Kamran Arshad}
is a Senior Member of the Institute of Electrical and Electronics Engineers (IEEE), currently holds the distinguished position of Dean of Research and Graduate Studies and Professor of Electrical Engineering at Ajman University in the United Arab Emirates. Dr. Arshad has played a pivotal role in leading numerous locally and internationally funded research projects that encompass the fields of cognitive radio, LTE/LTE-Advanced, 5G, optimization, and cognitive machine-to-machine communications. He has made significant contributions to several European and international large-scale research projects and has over 150 technical peer-reviewed articles published in esteemed journals and international conferences. He has been the recipient of three Best Paper Awards, one Best Research and Development Track Award, and has chaired technical sessions in several leading international conferences. Dr. Arshad holds the position of Associate Editor of the Journal on Wireless Communications and Networking (EURASIP).
\end{IEEEbiography}

\vskip -2\baselineskip plus -1fil

\begin{IEEEbiography}[{\includegraphics[width=1in,height=1.25in,clip,keepaspectratio]{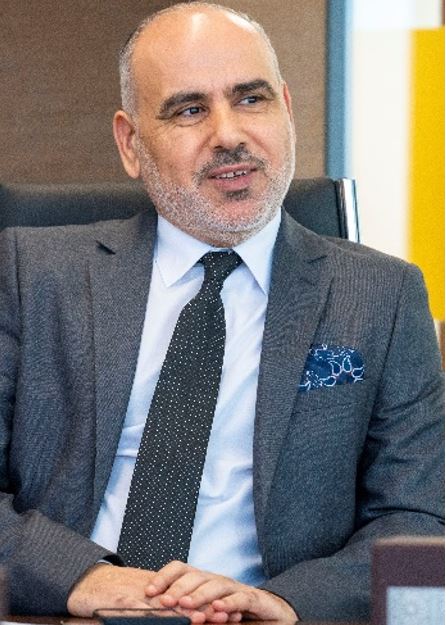}}]{Khaled Assaleh} is currently the Vice Chancellor for Academic Affairs and a Professor of Electrical Engineering at Ajman University. From 2002 through 2017, he was with the American University of Sharjah (AUS). Prior to joining AUS, Khaled had a 9-year R\&D career in the Telecom Industry in the USA with Rutgers, Motorola and Rockwell/Skyworks. He earned his Ph.D. in Electrical Engineering from Rutgers, the State University of New Jersey in 1993; M.S in Electronic Engineering from Monmouth University, New Jersey in 1990; and a B.Sc. in Electrical Engineering from the University of Jordan in 1988. He holds 11 US patents and has published over 130 articles on signal/image processing and machine learning and their applications. He has served on organization committees of several international conferences including ICIP, ISSPA, ICCSPA, MECBME and ISMA. He has also served as a guest editor for several special issues of journals. Dr. Assaleh is a senior member of the IEEE. His research interests include bio-signal processing, biometrics, speech and image processing, and pattern recognition.
\end{IEEEbiography}

\vskip -2\baselineskip plus -1fil

\begin{IEEEbiography}[{\includegraphics[width=1in,height=1.25in,clip,keepaspectratio]{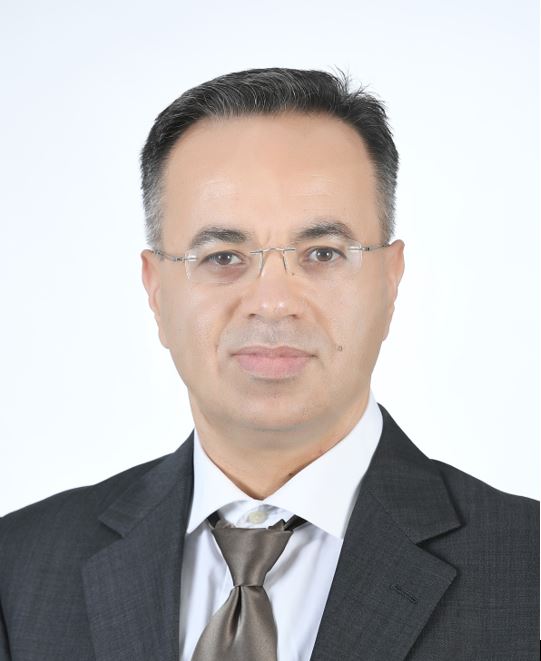}}]{Sami Muhaidat} received the Ph.D. degree in Electrical and Computer Engineering from the University of Waterloo, Waterloo, Ontario, in 2006. From 2007 to 2008, he was an NSERC Postdoctoral Fellow in the Department of Electrical and Computer Engineering, University of Toronto, Canada. From 2008-2012, he was an Assistant Professor in the School of Engineering Science, Simon Fraser University, BC, Canada. He is currently a Professor at Khalifa University and an Adjunct Professor with Carleton University, Ontario, Canada.  Sami's research interests focus on advanced digital signal processing techniques for wireless communications, intelligent surfaces, MIMO, optical communications, massive multiple access techniques, backscatter communications, and machine learning for communications.  He is currently an Area Editor of the IEEE Transactions on Communications, a Guest Editor of the IEEE Network “Native Artificial Intelligence in Integrated Terrestrial and Non-Terrestrial Networks in 6G” special issue, and a Guest Editor of the IEEE OJVT “Recent Advances in Security and Privacy for 6G Networks” special issue.  He served as a Senior Editor and Editor of the IEEE Communications Letters, an Editor of the IEEE Transactions on Communications, and an Associate Editor of the IEEE Transactions on Vehicular Technology.  
\end{IEEEbiography}

\vskip -2\baselineskip plus -1fil

\begin{IEEEbiography}[{\includegraphics[width=1in,height=1.25in,clip,keepaspectratio]{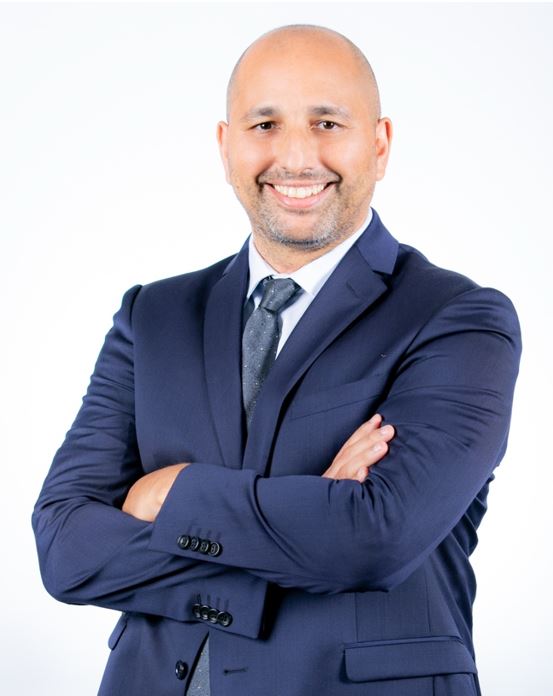}}]{Merouane Debbah} received the M.Sc. and Ph.D. degrees from the Ecole Normale Supérieure Paris-Saclay, France. He was with Motorola Labs, Saclay, France, from 1999 to 2002, and also with the Vienna Research Center for Telecommunications, Vienna, Austria, until 2003. From 2003 to 2007, he was an Assistant Professor with the Mobile Communications Department, Institut Eurecom, Sophia Antipolis, France. In 2007, he was appointed Full Professor at CentraleSupelec, Gif-sur-Yvette, France. From 2007 to 2014, he was the Director of the Alcatel-Lucent Chair on Flexible Radio. From 2014 to 2021, he was Vice-President of the Huawei France Research Center. He was jointly the director of the Mathematical and Algorithmic Sciences Lab as well as the director of the Lagrange Mathematical and Computing Research Center. Since 2021, he is Chief Research Officer at the Technology Innovation Institute in Abu Dhabi. He leads jointly the AI and Telecommunication centers. He has managed 8 EU projects and more than 24 national and international projects. He is an IEEE Fellow, a WWRF Fellow, a Eurasip Fellow, an Institut Louis Bachelier Fellow and a Membre émérite SEE. He was a recipient of the ERC Grant (Advanced Mathematical Tools for Complex Network Engineering) from 2012 to 2017.
\end{IEEEbiography}

\vskip -2\baselineskip plus -1fil

\begin{IEEEbiography}[{\includegraphics[width=1in,height=1.25in,clip,keepaspectratio]{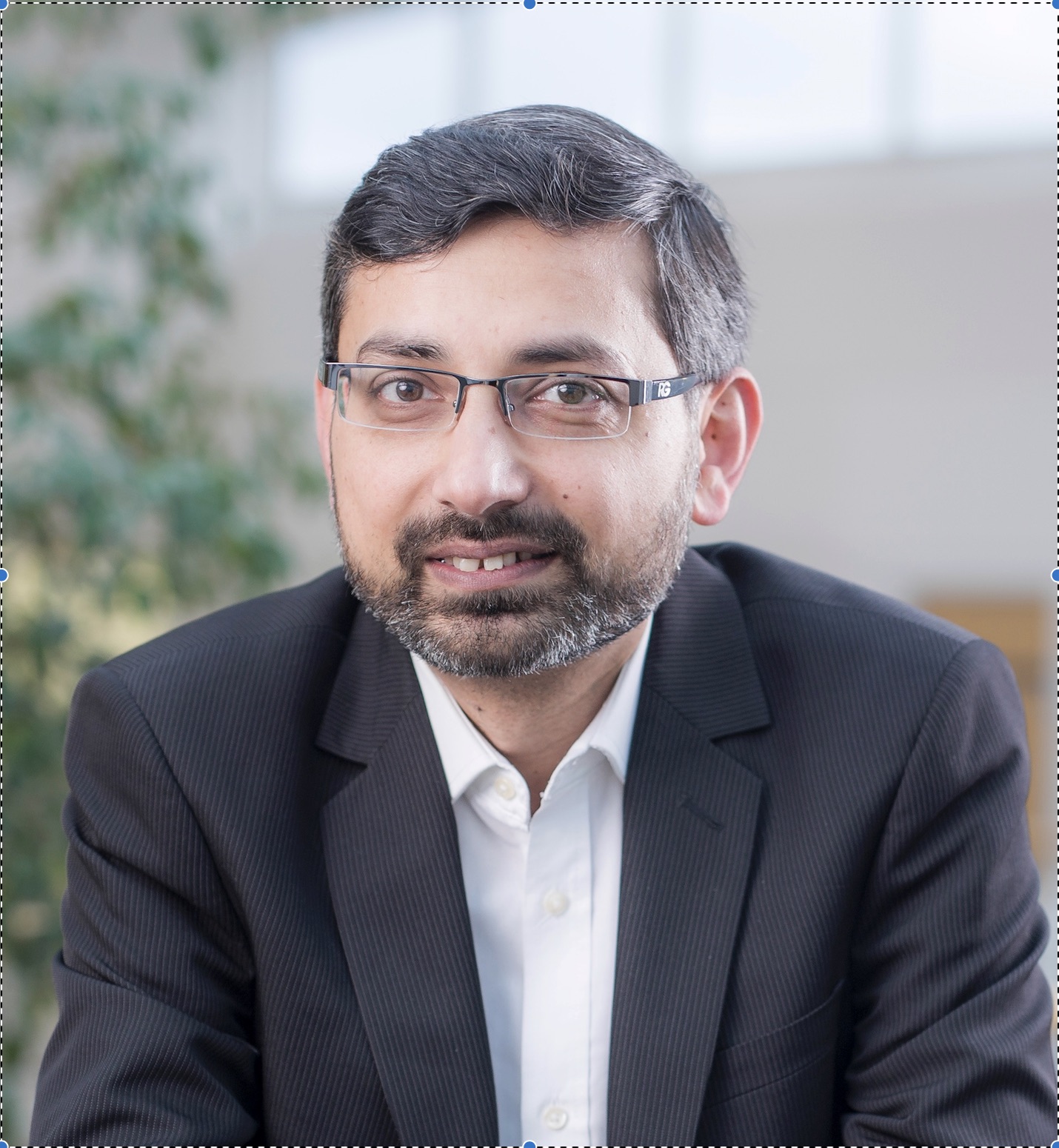}}]{Muhammad Ali Imran}
received the M.Sc. (Hons.) and Ph.D. degrees from Imperial College London, U.K., in 2002 and 2007, respectively. He is currently the Dean of University of Glasgow UESTC,  Head of Autonomous Systems and Connectivity Division and a Professor of communication systems with the James Watt School of Engineering, University of Glasgow, U.K. He is an Affiliate Professor at the University of Oklahoma, USA and 5G Innovation Centre, University of Surrey, U.K. He is leading research at the University of Glasgow for Scotland 5G Centre. He has over 20 years of combined academic and industry experience with several leading roles in multi-million pounds funded projects, working primarily in the research areas of cellular communication systems. He has been awarded 10 patents, has authored/co-authored over 500 journals and conference publications, and has supervised more than 50 successful Ph.D. graduates. He has an Award of Excellence in recognition of his academic achievements, conferred by the President of Pakistan. He was awarded the IEEE Comsoc’s Fred Ellersick Award 2014, the FEPS Learning and Teaching Award 2014, and the Sentinel of Science Award 2016. He is a Fellow of Royal Society of Edinburgh, Fellow of the Institution of Engineering and Technology, Fellow of Royal Society of Arts and a Senior Fellow of the Higher Education Academy. He is speciality chief editor for IoT section of Frontiers in Communications and Networks and an Associate Editor for IEEE Transactions on Communications, and previously served in editorial roles for the IEEE Communications Letters, the IEEE Access, and the IET Communications Journals. He is a Senior Member of IEEE.
\end{IEEEbiography}

\end{document}